\documentclass[fleqn,usenatbib]{mnras}

\usepackage{newtxtext,newtxmath}

\usepackage[T1]{fontenc}
\usepackage{cancel}
\usepackage{xcolor}

\DeclareRobustCommand{\VAN}[3]{#2}
\let\VANthebibliography\thebibliography
\def\thebibliography{\DeclareRobustCommand{\VAN}[3]{##3}\VANthebibliography}
\def\gtorder{\mathrel{\raise.3ex\hbox{$>$}\mkern-14mu
    \lower0.6ex\hbox{$\sim$}}}
\def\ltorder{\mathrel{\raise.3ex\hbox{$<$}\mkern-14mu
    \lower0.6ex\hbox{$\sim$}}}

\usepackage{graphicx}	
\usepackage{amsmath}	

\title[Vertical Buckling in Galactic Bars]{The Origin of Buckling Instability in Galactic Bars:\\ Searching for the Scapegoat} 

\author[Xingchen Li et al.]{
Xingchen Li,$^{1}$\thanks{E-mail: xingchen.li@uky.edu}
Isaac Shlosman,$^{1,2}$\thanks{E-mail: isaac.shlosman@uky.edu}
Daniel Pfenniger$^{3}$
and Clayton Heller$^{4}$
\\
$^{1}$Department of Physics \& Astronomy, University of Kentucky, Lexington, KY 40506-0055, USA\\
$^{2}$Theoretical Astrophysics, Graduate School of Sciences, Osaka University, Osaka, Japan\\
$^{3}$University of Geneva, Geneva Observatory, ch. Pegasi 51, 1290 Versoix, Switzerland\\
$^{4}$Department of Physics \& Astronomy, Georgia Southern University, Statesboro, GA 30460, USA
}

\date{Accepted XXX. Received YYY; in original form ZZZ}

\pubyear{2022}

\begin{document}
\label{firstpage}
\pagerange{\pageref{firstpage}--\pageref{lastpage}}
\maketitle

\begin{abstract}
The buckling process in stellar bars is full of ambiguities and unsolved issues. We analyze the origin of the buckling instability in stellar bars by means of high-resolution $N$-body simulations. Previous studies have promoted the nonresonant firehose instability to be responsible for the vertical buckling. We have analyzed the buckling process in terms of the resonant excitation of stellar orbits in the bar region, which pumps energy into vertical stellar oscillations. We find that (1) the buckling is associated with an abrupt increase in the central mass concentration and triggers velocities along the bar and along its rotation axis. The velocity field projected on one of the main axes forms circulation cells and increases vorticity there, which are absent in classical firehose instability; (2) The bending amplitude is nonlinear when measured by isodensity contours or curvature of the Laplace plane, which has a substantial effect on the stellar motions in the bar midplane; (3) In the linear description, the planar and vertical 2:1 resonances appear only with the buckling and quickly reach the overlapping phase, thus supporting the energy transfer from horizontal to vertical motions; (4)  Using  nonlinear orbit analysis, we analyze the stellar oscillations along the bar and along the rotation axis and find that stars cross the vertical 2:1 resonance {\it simultaneously} with the buckling. The overlapping planar and vertical 2:1 resonances trapping more than 25\% of the bar particles provide the `smoking gun'  pointing to a close relationship between the bending of stellar orbits and the resonant action --- these particles provide the necessary ingredient assuring the cohesive (i.e., collective) response in the growing vertical asymmetry. Therefore, we conclude that resonant excitation is important in triggering the buckling instability, and the contribution from the nonresonant firehose instability should be reevaluated. Finally, we discuss some observational implications of buckling on the kinematics in face-on and inclined galactic disks.     
\end{abstract}




\section{Introduction}
\label{sec:intro} 

Galactic bars are ubiquitous --- majority of disk galaxies are barred \citep[e.g.,][]{sell93,marti95,kna00,gros04,mene07,erwin18}. Modeling of pure stellar bars is in essence an $N$-body problem which has been analyzed for the last half a century. Yet, we still do not have a clear understanding of the bar formation and their dynamics. Among these basic problems, the issue of the vertical buckling instability in bars stands out. 

Stellar bars experience a spontaneous breaking of their vertical symmetry after reaching their maximal strength --- the so-called buckling or bending instability. The final product of this buckling, after the vertical symmetry has been restored, is the boxy/peanut shaped bulges. The reasons for the origin of this instability are not completely understood, and this paper deals with the physical processes which potentially govern this instability. In particular, we focus on the firehose instability as an explanation to buckling and test the role of the orbital resonances in the buckling.

We do not deal here with the formation of the boxy/peanut bulges and only comment about their appearance in passing. Such a bulge profile has been observed in about 50\% of edge-on disk galaxies \citep[e.g.,][]{luti00}. This frequency was found to depend on the stellar mass of galaxies and other galaxy parameters \citep{erwin17}, and such bulges have been detected in the range of $z = 0-1$ \citep{kruk19}. The presence of the boxy/peanut shaped bulges is based on the existence of a specific family of periodic 3-D orbits, and should not be confused with the buckling instability itself. These orbits exist before and after buckling, and can be populated by a variety of processes, such as vertical buckling, resonant scattering of stars trapped in the bar, galaxy interactions, etc. 

Indeed, theoretical explanations of the formation of the boxy/peanut-shaped bulges have been proposed as due to the resonant action of stellar orbits with the bar \citep[][]{comb90}, and alternatively by the firehose instability \citep[][see \citet{binn08} for review]{toom66,raha91}. Note that \citet{pfen91} skipped discussing the break of the vertical symmetry, their analysis of stellar orbits aimed only on the final product of vertical instability.  However, the role of the resonances in triggering the buckling itself has been never analyzed. This paper focuses on this issue.

The vertical buckling (or bending) instability in a non-rotating planar stellar system, i.e., horizontally homogeneous slabs and layers, has been analyzed theoretically for more than half a century \citep[e.g.,][]{toom66,kuls70,kuls71,mark71,araki85}. However, it is unclear whether such simplified models are relevant for inhomogeneous and rotating galactic bars. In numerical simulations of barred disk galaxies, buckling has been first noticed by \citet{friedli90}.  The instability is dynamical and results in the vertical thickening of the bar, which acquires a characteristic peanut/boxy profile in the central region.  Moreover, this buckling can be recurrent \citep[][see \citet{shlo13} for review]{marti06}. 

This paper is structured as follows. Section\,\ref{sec:about} introduces the firehose instability and difficulties associated with identifying it with the buckling instability. Section\,\ref{sec:numerics} exhibits our numerical methods and initial conditions. Our results are presented in section\,\ref{sec:results}, and are followed by comparison of the firehose and resonance triggering of the buckling instability, and by conclusions.

\section{Buckling: so what is it about?}
\label{sec:about} 

Buckling out of the plane requires an increase in the energy of vertical motions in the disk. Hence, a source of energy must be found in the disk energy. In addition, a mechanism which facilitates the energy transfer must exist. This mechanism can be based on the action of vertical resonances or be a nonresonant one in nature. In this section, we present an overview of the firehose instability and its alternative --- orbital excitation by the vertical resonances. The analysis of our results will be presented in section\,\ref{sec:discussion}.

If the classical firehose instability is the underlying cause of the buckling, the instability is a kinematic one and does not involve any resonances --- it has been analyzed assuming a razor-thin stellar disk or a thin slab, independently by \citet{toom66} and by \citet{kuls71}, who pointed out the similarity between the firehose instability in the stellar slabs or layers and previously studied firehose instability in plasma\footnote{In the astronomical literature, the firehose instability is sometimes confused with the Kelvin-Helmholtz instability. \citet{toom66} actually only evokes analogies between these instabilities.  The two instabilities actually imply different physical conditions: the former occurs only in a collisionless fluid because its velocity dispersion tensor is strongly anisotropic, while the latter occurs in a collisional fluid (thus with an isotropic velocity dispersion) because the velocity gradient between stratified layers is non-zero.}. 

The linear analysis of this instability shows that the slab is unstable for sufficiently small perturbation wavelengths, specifically for wavelengths smaller than  the Jeans length, $\uplambda_{J}\sim \upsigma_{R}^2/G\Sigma$, where $\upsigma_{R}$ is the radial dispersion velocity, $\Sigma$ is the surface density of the slab, and $R$ is the cylindrical radius. The linear dispersion relation for an infinitely-thin sheet or thin slab with small bending amplitude\footnote{Small bending amplitude here assumes also that the bending slope is small, meaning that it does not affect the radial motion of stars.} can be written in the form \citep{toom66} as
\begin{equation}
\omega^2 = 2\pi G\Sigma k -\sigma_{R}^2 k^2,    
\end{equation}
where $k$ is the instability wavenumber. In this picture, the instability occurs when the centrifugal force acting on particles which travel through the buckled layer exceeds the restoring gravitational force.  

However, this analytical treatment ignores the nonzero vertical thickness of the bar, $h$, and its vertical stellar dispersion velocities $\upsigma_{z}$, which can be related via $h\sim \upsigma_{z}^2/G\Sigma$. For the 3-D stellar disk, $\upsigma_{z} < \upsigma_{R}$, and hence, the firehose instability occurs in the wavelength range of $\upsigma_{z}^2/G\Sigma < \uplambda < \upsigma_{R}^2/G\Sigma$, or $\upsigma_{z}^2/\upsigma_{R}^2 < 1$. Therefore, an increase in the ratio $\upsigma_{z}^2/\upsigma_{R}^2$ can stabilize the disk against the vertical buckling. Of course, the fate of this instability for $\uplambda/h < 1$ cannot be resolved analytically and must be treated numerically.

The physics of the firehose instability involves competition between the disk vertical gravity and the (vertical) centrifugal force acting on the stars moving along the bar. The disk gravity acts as a stabilizing forces while the stellar motion along the bar destabilizes the motion. If the disk is embedded in the dark matter (DM) halo, its gravity normal to the disk adds to the restoring force. Finite thickness disks reduce the range of unstable wavelengths by stabilizing the shortest wavelengths \citep{toom66,polya77,araki85}. But \citet{merr94} argued instead that thick disks with $\uplambda < h$ are stabilized by an out-of-phase response of stars that encounter the bend with the frequency $\nu > \nu_{z}$ (i.e., larger than the frequency of vertical oscillations). 

An infinite slab or layer, when unstable, can buckle anywhere, as there is no preferred position for the instability. So why do bars always buckle at the rotation axis? Within the context of the firehose instability, the answer lies in that radial velocity dispersion in stellar bars always peaks at the center, and if the restoring gravity force is weaker than the centrifugal force (which also depends on the unstable wavelength), they will buckle at the center.  

The firehose instability analysis for the razor-thin bar or disk removes all the vertical resonances. Furthermore, treating the bar or the disk as infinite razor-thin and nonrotating systems overestimates the stabilizing action of gravity by about a factor of $\sim 2$ \citep{merr94}. As these authors have shown, the introduction of an arbitrary radial density and velocity dispersion profile in the system can have a similar effect as the reduction of stabilizing gravity. Under the above conditions, the buckling instability can occur even for wavelengths larger than the Jeans wavelengths. 

However, we note that fundamental differences exist between the stability of disks and bars. First, disks can be radially cold systems, i.e., their radial velocity dispersion can be zero, meaning that stars move on circular orbits. In contrast, bars are by definition associated with large radial velocities. Stellar bars are always radially hot, as the fundamental family of stellar orbits in bars are aligned with the bar. This also explains why the bar instability always precedes the buckling. 

Second, bars are finite systems and their radial dispersion velocities peak at the center, irrespective of the density distribution. As a result, bars, if buckle, will always buckle at the center, resulting in the vertical Fourier $m=0$ mode, unlike disks. In this respect, the buckling instability, even for the non-uniform density, discussed in \citet{merr94} has very little application to stellar bars. Specifically, their speculation that stellar bars can be destroyed by the buckling has been disputed \citep[][]{marti04}, and no numerical model supporting such a dramatic outcome has been presented so far.

Hence, stability of unbarred disks to buckling differs from that of the stellar bars. In this work, we focus strictly on the bar dynamics --- a finite-size inhomogeneous system with anisotropic velocity dispersion, and hence avoid all the issues associated with stability of stellar disks to bending. 

That stellar orbits of barred disk galaxies can be destabilized out of the plane along the $z$-axis and populate the 3-D orbits has been recognized early \citep[e.g.,][]{binn81,comb81,mart81,cont82,pfen84}. Numerical search for 3-D periodic orbits in barred galaxies has found the extension of the main planar orbits forming the backbone of the bar, the so-called banana (BAN) and anti-banana (ABAN) shaped orbits \citep[e.g.,][]{comb90,pfen91,marti06}.  But the nonlinear mode evolution can be only treated using numerical modeling.  In this paper we focus on some aspects of this nonlinear evolution of this instability. We have performed a fully 3-D orbital analysis for our models and present the most important orbits in the Appendix.
 
Recent work by \citet{sell20} describes three numerical experiments. The first experiment reiterates the arguments that the vertical thickening of the stellar bar is a result of the firehose instability. It is not clear, however, what is the mechanism of the collective response of stellar orbits during the buckling process. The second and third experiments have introduced the massive bulge and imposed the vertical symmetry in the bar, suppressing the buckling, and concluding that the vertical thickening of the bar results from the action of the vertical 2:1 resonance. Thus confirming the results of \citet{friedli90}. 
 
The alternative explanation of the formation of peanut/boxy bulges promoting the resonant action has circumvented the buckling phenomenon \citep[e.g.,][]{comb90}. It argued that the rotational energy in the disk plane can be pumped into vertical oscillations of bar stars, mostly by the action of the vertical inner Lindblad resonance (vILR). This action results in the formation of the peanut/boxy shaped bulges and consequently in the overall increase in the vertical thickness of the stellar bar. 

Note that only the appearance of boxy/peanut shaped bulges has been detected in \citet{comb90}, while buckling itself has been overlooked, probably due to the lower available resolution. To demonstrate the crucial role of resonances in forming the peanut/boxy-shaped bulges and that these bulges can form without the buckling instability, \citet{friedli90} have suppressed buckling by artificially enforcing $z$-symmetry. The peanut/boxy bulges formed in the absence of buckling, though over a secular instead of a dynamical timescale. The corresponding model with buckling have shown the formation of these bulges as well. Thus bucking is a sufficient but not necessary condition for the occurrence of peanut/boxy-shaped bars.

Nonetheless, the classical firehose instability scenario does not predict the complex conditions in a finite inhomogeneous and rotating bar, while the vertical resonance scenario lacks an explanation about the collective response of an ensemble of particles. So arguments have been used in favor of the firehose instability as the trigger of buckling in stellar bars \citep[e.g.,][]{sell20}. And no detailed analysis has been presented so far in favor of the resonance trigger. A decisive argument in favor of either of these is elusive, and the problem resembles that of the chicken and egg. It is not clear which is the cause and which is the consequence. We attempt to provide  here some clarity on this issue. 
 
The buckling instability of the stellar bars has a number of features that can be potentially observable. The most obvious one is, as stated above, the appearance of the peanut/boxy bulges. We have already mentioned that about 50\% of the edge-on disk galaxies display such bulges in the contemporary universe \citep[e.g.,][]{luti00}, and up to $z\sim 1$ \citep{kruk19}.

Buckling in the gas-rich barred galactic disks has a sharply decreased amplitude \citep{bere98,debat06,bere07,villa10}, which is especially relevant for the high-redshift galaxies \citep{bi22}. The DM halo mass concentration affects the timing of the buckling \citep{villa09}.

\citet{lokas19a} has analyzed the buckling in four interacting modeled galaxies and produced velocity fields in these objects. However, both the distortion modes and the velocity field include the bars and tidally-distorted galactic disks. It is very difficult to disentangle the effects of the stellar bar and tidally-perturbed disk under these circumstances. Both morphological and kinematical features are, therefore, difficult to interpret. The only unequivocal signature of the bar during the maximal buckling amplitude displays the face-on disk velocity averaged along the $z$-axis --- it shows the double dumbbell, one with positive and one with negative velocities, reflecting formation of the peanut bulge. This feature agrees with our inner circulation cell (Figs.\,\ref{fig:vx_accum} and \ref{fig:vz_accum}), both in position and the velocity value. However, the conclusions are more straightforward for the case of an isolated galaxy \citep{lokas19b}.

The same kinematic signature of double dumbbell or quadrupolar pattern has been observed in numerical simulations of \citet{xiang21}. Moreover, analyzing the sample of galaxies from the MaNGA survey \citep{bundy15}, they have been able to select five candidate galaxies with this kinematic feature, hence caught in the process of buckling. Further claims of observing stellar bars in the process of buckling have been promoted for two local galaxies, NGC\,3227 and NGC\,4569, based on the surface brightness profiles \citep{erwin16}.

\section{Numerics}
\label{sec:numerics} 

We use the $N$-body part of the mesh-free hydrodynamics code \textsc{GIZMO} \citep{hopk17}. The models of stellar disk galaxies embedded in spherical DM halos have been constructed with different DM densities and a sequence of cosmological spins $\uplambda$. Here we discuss the case of a nonrotating halo, while $\uplambda > 0$ models are analyzed in the followup paper (Li et al., in prep.). 

The units of mass, length, and velocity have been chosen as $10^{10}\,\mathrm{M_{\odot}}$, $1\,\mathrm{kpc}$, and $1\, \mathrm{km\,s^{-1}}$, respectively. The resulting unit of time is $1\,\mathrm{Gyr}$. The number of DM halo particles has been taken as $N_\mathrm{DM}=7.2\times 10^6$, and stellar disk particles $N_\mathrm{S}=8 \times 10^5$, in order to have a similar mass-per-particle. The gravitational softening length for both DM and stellar particles is $\epsilon_{\mathrm{DM}} = \epsilon_{\mathrm{*}} = 25 \, \mathrm{pc}$. The opening angle $\theta$ of the tree code has been reduced from $0.7$ used in cosmological simulations to 0.4 for a better quality of force calculations. The models have been run for $10\,\mathrm{Gyr}$, with the angular momentum conservation within 0.6\% and energy conservation within 0.2\%. 


\subsection{Initial conditions}
\label{sec:ICs}

The initial conditions are those of an exponential stellar disk with a density profile given by
\begin{equation}
    \rho_{\mathrm{d}}(R, z) = \left( \frac{M_\mathrm{d}}{4 \pi R_0^2 z_0} \right) \, \exp\left(-\frac{R}{R_0}\right) \, \mathrm{sech}^2 \left( \frac{z}{z_0} \right),
    \label{eq:rho_d}
\end{equation}
where $R$ is the cylindrical radius, $M_\mathrm{d} = 6.3 \times 10^{10} \, \mathrm{M_{\odot}}$ is the mass of the disk , $R_0 = 2.85 \, \mathrm{kpc}$ is the disk radial scalelength, and $z_0 = 0.6 \, \mathrm{kpc}$ is the disk scaleheight. The disk is truncated at $6 R_{\mathrm{0}} \sim 17 \, \mathrm{kpc}$, i.e., at 98\% of its mass.

The initial DM component consists of a spherical halo with the density profile as a function of spherical radius $r$ described by \citet[][hereafter NFW]{nfw96},
\begin{equation}
    \rho_\mathrm{h} (r) = \frac{\rho_{\mathrm{s}} }{[(r + r_\mathrm{c}) / r_\mathrm{s}] (1 + r / r_\mathrm{s})^2} \, e^{- (r / r_\mathrm{t})^2},
\end{equation}
where $\rho_{\mathrm{s}}$ is a normalization parameter, $r_{\mathrm{c}} = 1.4 \, \mathrm{kpc}$ is the size of the flat density core, and $r_{\mathrm{s}} = 10 \, \mathrm{kpc}$ is a characteristic radius. We use a Gaussian cut-off radius $r_{\mathrm{t}} = 180$\,kpc to obtain the finite mass.  

We use an iteration method introduced by \citet{rodio06} to obtain the velocity distribution function for the DM, see also \citet{rodio09}, \citet{long14} and \citet{coll18}. For each iteration, we freeze the disk, then release the DM particles from the initial density distribution for $\sim 0.3\,\mathrm{Gyr}$. Next, we return the DM particles to their nearest unevolved particles in the initial distribution, with the new velocities. Typically, about $50-70$ iterations have been used, until the virial ratio and the velocity distribution of DM particles converge.

For the stellar disk, we use the epicycle approximation to determine the disk velocity. The stellar dispersion velocities are
\begin{equation}
    \sigma_{R} (R) = \sigma_{R,0} \, \exp \left( -\frac{R}{2R_0} \right)
\end{equation}
\begin{equation}
    \sigma_{z} (R) = \sigma_{z,0} \, \exp \left( -\frac{R}{2R_0} \right),
\end{equation}
where $\sigma_{z,0} = 120\,\mathrm{km/s}$, and $\sigma_{R,0}$ is determined by setting the minimal Toomre parameter \citep[e.g.,][]{binn08}, to $Q = 1.5$, at $\sim 2R_0$.

\subsection{The stellar bar characteristics}
\label{sec:a2}

In all models, stellar bars have developed from the initially axisymmetric mass distribution. We quantify the bar strength by the Fourier components of the surface density. For $m$ mode, the Fourier amplitude is $\sqrt {a_{m}^2 + b_{m}^2}$, where 
\begin{equation}
    a_{ m}(R) = \frac{1}{\pi} \int_{0}^{2\pi} \Sigma(R, \theta) \, \cos(m\theta) \, \mathrm{d} \theta, \; \; \; \; \; m=0, 1, 2, \ldots ,
\end{equation}
\begin{equation}
    b_{ m}(R) = \frac{1}{\pi} \int_{0}^{2\pi}\Sigma(R, \theta) \, \sin(m\theta) \, \mathrm{d} \theta, \; \; \; \; \; m=0, 1, 2, \ldots  ,
\end{equation}    
and $\Sigma(R,  \theta)$ is the surface stellar density. To quantify the bar strength, we use the normalized $A_2$ amplitude which is defined as 
\begin{equation}
    \frac{A_2}{A_0}=\frac{ \int_{0}^{R_{\mathrm{max}}} \sqrt {a_{2}^2 + b_{2}^2} \, \mathrm{d} R}{ \int_{0}^{R_{\mathrm{max}}} a_0 \, \mathrm{d} R }.
    \label{eq:barA2}
\end{equation}
We choose the upper limit of integration, $R_{\mathrm{max}}$, as the radius which contains $98\%$ of the disk mass at a given time.  

Similarly, we measure the vertical buckling strength, $A_{1z}$, i.e. the vertical asymmetry, by calculating the the $m=1$ Fourier amplitude in the $xz$-plane, where the major axis of the bar is aligned with the $x$-axis, and the rotation axis is along $z$, as follows:
\begin{equation}
    \frac{A_{1z}}{A_0}=\frac{ \int_{-x_0}^{x_0} \sqrt {a_{1}^2 + b_{1}^2} \, \mathrm{d} x}{ \int_{-x_0}^{x_0}a_0 \, \mathrm{d} x }.
    \label{eq:barA1z}
\end{equation}
The integral is over the region $|x| < 12$ kpc, $|y| < 3$ kpc, $|z| < 5$ kpc.

Lastly, the angle of the bar, $\phi_{\rm bar}$, is obtained from

\begin{equation}
\phi_{\rm bar} = \frac{1}{2} \arctan\left(\frac{b_2}{a_2}\right),    
\end{equation}
and the bar pattern speed $\Omega_{\mathrm{bar}}$ is calculated by taking a discrete derivative of $\phi_{\mathrm{bar}}$ with respect to time.

\section{Results}
\label{sec:results}

\begin{figure}
\center 
	\includegraphics[width=0.48\textwidth]{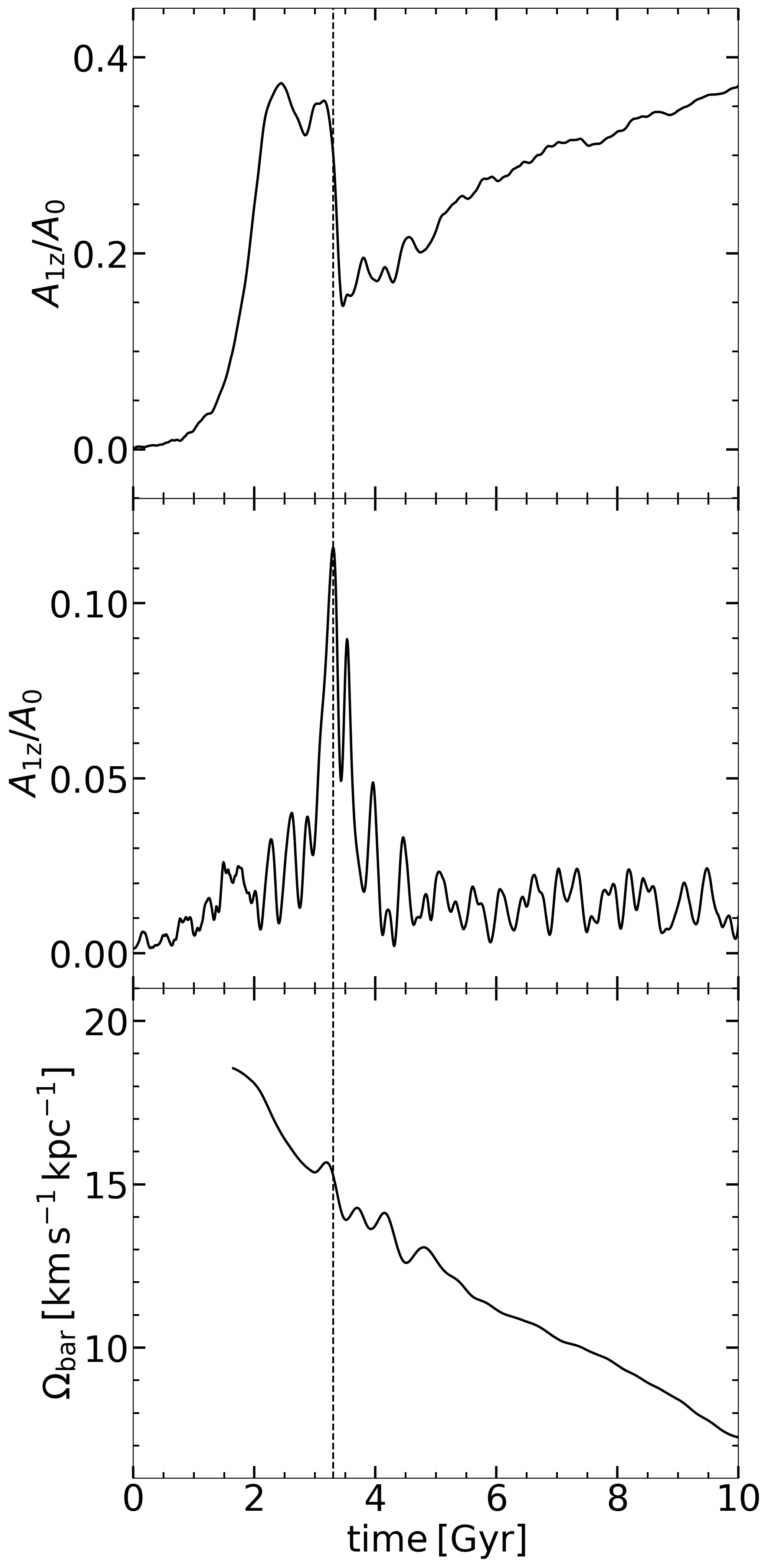}
    \caption{Evolution of the normalized Fourier amplitude $A_{\rm{2}}$, the buckling amplitude $A_{1z}$ of the stellar bar, and the bar pattern speed $\Omega_{\rm bar}$,  for the standard model with a non-rotating halo. The vertical dashed line marks the time of the maximum buckling at $t=3.3 \, \mathrm{Gyr}$}
    \label{fig:a2etc}
    \end{figure}

\begin{figure}
\center 
	\includegraphics[width=0.46\textwidth]{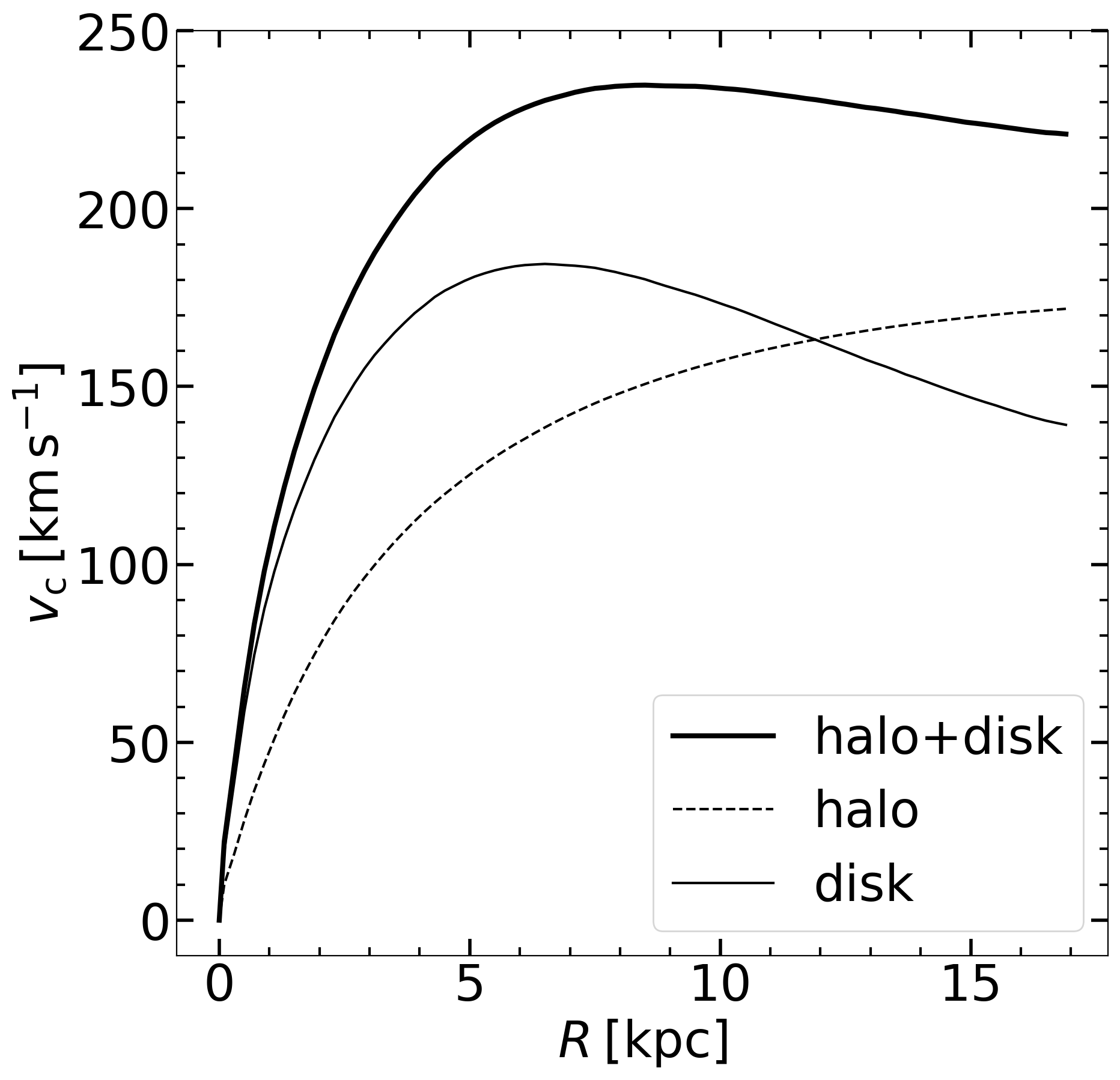}
    \caption{Circular velocity of the stellar disk, DM halo and their total, as a function of radius $R$ at $t=0$. }
    \label{fig:vc}
    \end{figure}

\begin{figure}
\center 
	\includegraphics[width=0.46\textwidth]{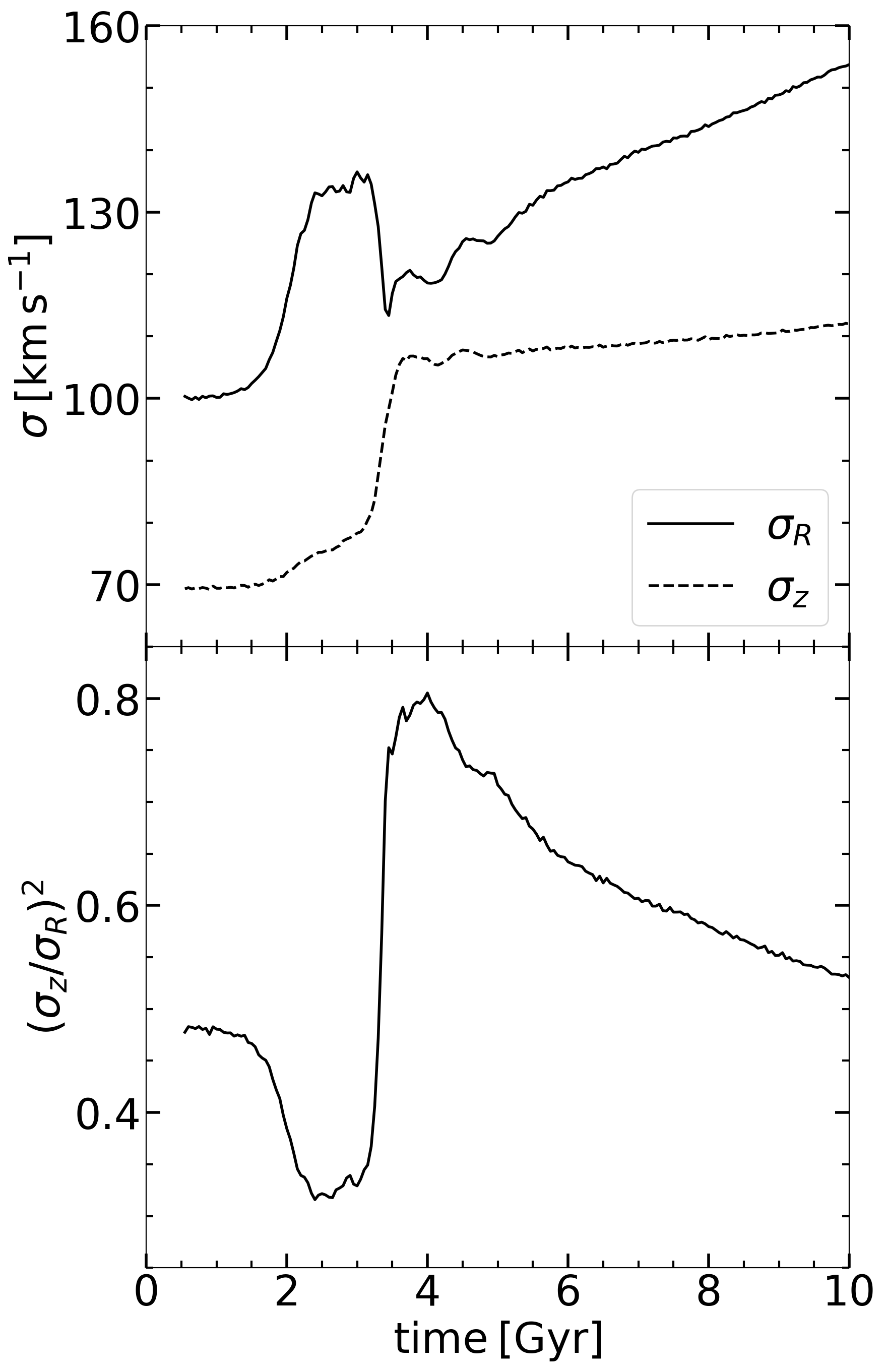}
    \caption{Evolution of stellar velocity dispersions in the bar region, i.e., $|x| < 4$\,kpc, $|y| < 2$\,kpc, $|z| < 3$\,kpc: radial $\sigma_{R}$, vertical $\sigma_{z}$, and their ratios $\sigma_{z}^2/\sigma_{R}^2$ for the stellar disk.}
    \label{fig:sigmas}
    \end{figure}

We focus on the vertical buckling of stellar bars in order to analyze the triggers of this instability, and its evolution well past its linear stage. Our goal is to deepen our understanding of this instability by analyzing evolution of the disk-halo system. 

\subsection{Evolution of the standard model}
\label{sec:standard}

As a first step, we have run the standard model of a bar development in a stellar disk embedded in DM halo for 10\,Gyr and analyzed it. Figure\,\ref{fig:a2etc} displays the evolution of the stellar bar amplitude, $A_2$, the bar pattern speed $\Omega_{\rm bar}$, and the vertical buckling amplitude, $A_{1z}$ --- all integrated over the bar length. This is a regular evolution of a stellar bar in an isolated nonrotating DM halo and is shown only to confirm that this is a typical system. Note, the buckling and the subsequent sharp decrease in $A_2$ happen around $t\sim 3.3$\,Gyr, which is followed by the renewed growth of the bar.

We also show the circular velocities of the stellar disk, DM halo, and their total contribution, $v_{\rm c},$ at $t=0$ in Figure\,\ref{fig:vc}, as well as  evolution of the dispersion velocities, $\sigma_{R}$ and $\sigma_{z}$, and their ratios $\sigma_{z}^2/\sigma_{R}^2$ (Fig.\,\ref{fig:sigmas}). Initially, the disk $v_{\rm c}$ dominates inside the central $\sim 12$\,kpc. The $\sigma_{R}$ evolution generally follows that of the $A_2$, while $\sigma_{z}$ is monotonically increasing, with a sharp increase during the buckling process. Hence, $\sigma_{z}^2/\sigma_{R}^2$ decreases from $\sim 0.48$ at $t=0$ to $\sim 0.32$. Then increases sharply to $\sim 0.8$ during buckling, when the stellar bar shortens and thickens vertically with the break of the vertical symmetry. Subsequently, this ratio declines over time to $\sim 0.55$ at the end of the run. The boxy/peanut-shaped bulge appears after buckling.

\begin{figure}
\center 
	\includegraphics[width=0.46\textwidth]{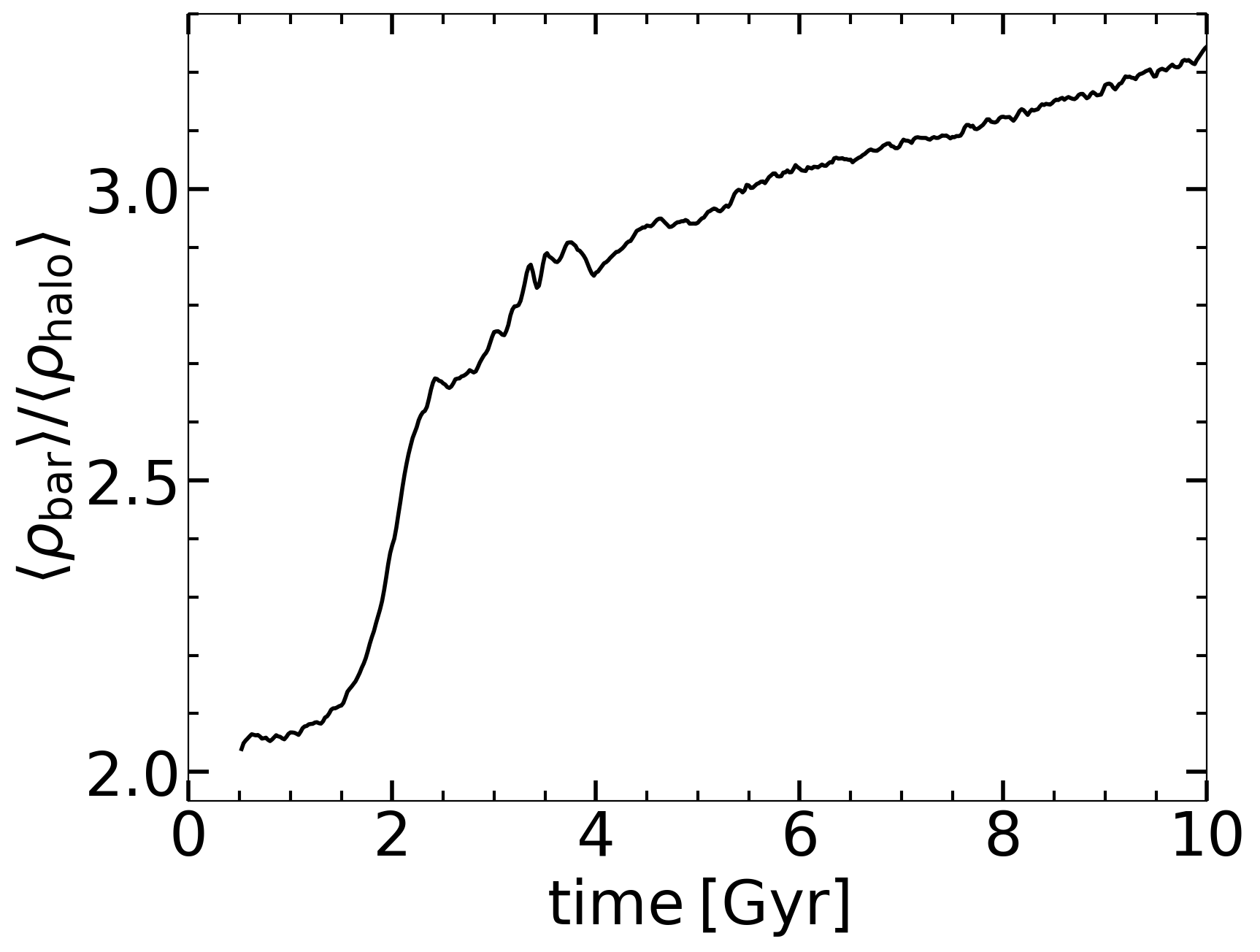}
    \caption{Evolution of the ratio of the stellar bar-to-DM halo average densities, $\langle\rho_{\rm bar}\rangle/\langle\rho_{\rm halo}\rangle$, for the standard model.  The density has been averaged over $|x| < 4$\,kpc along the bar, perpendicular to the bar $|y| < 2$\,kpc, and the vertical slice $|z| < 3$\,kpc.  }
    \label{fig:surf_dens}
    \end{figure}

As the bar develops in the initially axisymmetric disk, we follow the evolution of the ratio of stellar density in the bar to that of the DM (Fig.\,\ref{fig:surf_dens}). This ratio has been averaged over  $|x| < 4$\,kpc, $|y| < 2$\,kpc, and $|z| < 3$\,kpc, where the $x$-axis is aligned with the bar, and the stellar and DM densities have been integrated vertically along the $z$-axis and along the $y$-axis withing the above limits. The ratio grows from the initial values of 2.05 to about 2.7, when the bar reaches its maximal strength, then continues its growth to 2.9, when the bar experiences vertical buckling at around $t\sim 3.3$\,Gyr, saturates for a short time and continues its monotonic growth thereafter. Hence, the averaged stellar density within the inner region of the bar has increased substantially, to more than a factor of 3 over the averaged DM density --- an increase by about 50\%. The reason for this increase lies in the continuous trapping of disk orbits by the bar. These orbits become more and more radial as they lose their angular momentum to the outer disk and the DM.   

\begin{figure}
\center 
	\includegraphics[width=0.46\textwidth]{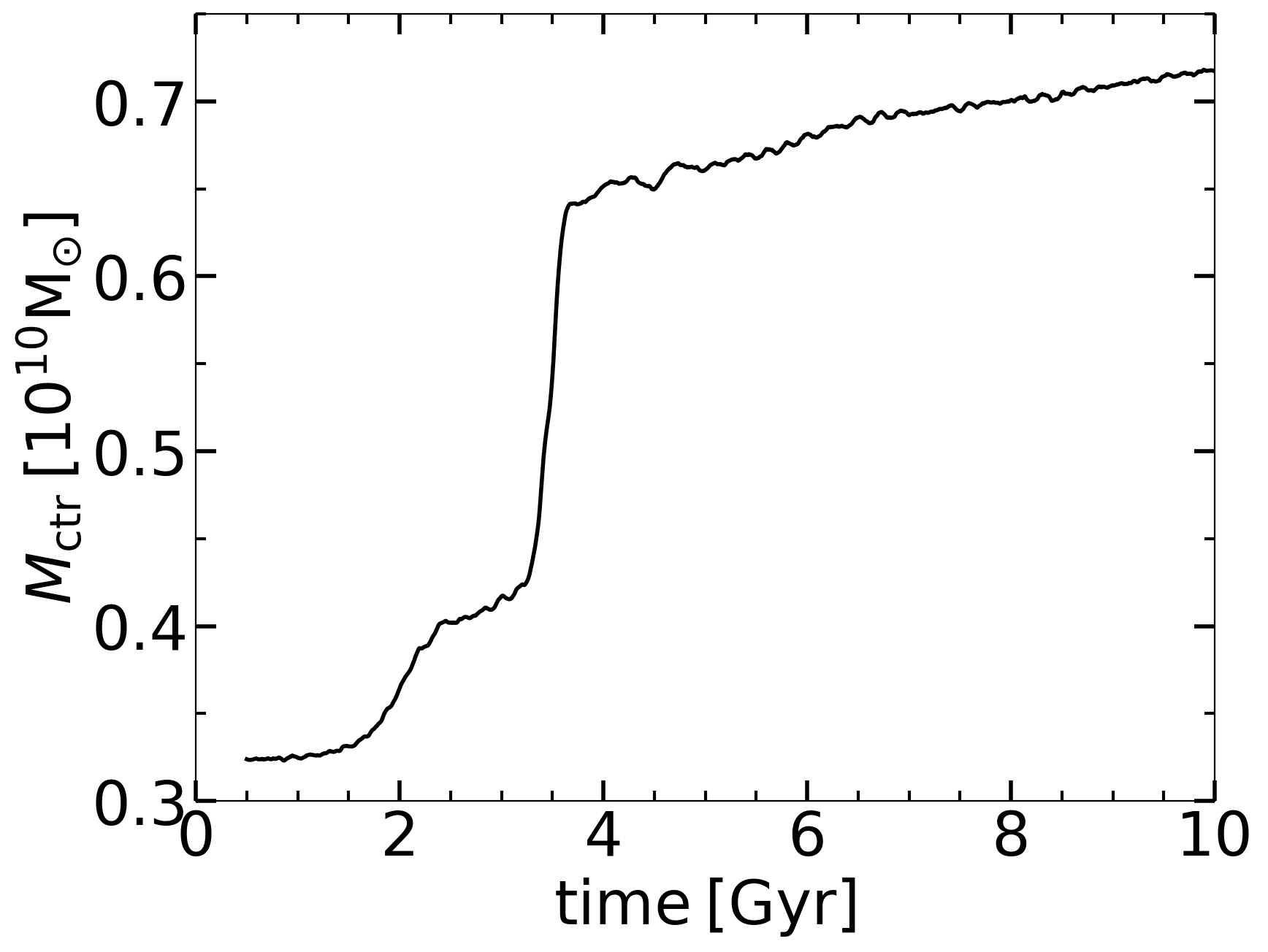}
    \caption{Evolution of the mass accumulation in the central region of the stellar bar in the standard model. The mass has been calculated within $|x| < 0.5$\,kpc along the bar, $|y| < 2$\,kpc perpendicular to the bar, and the vertical slice $|z| < 3$\,kpc. }
    \label{fig:mass_accum}
    \end{figure}

Such an increase in the average density in the central region of the stellar bar must invoke a mass inflow, and we analyze the radial mass redistribution in the central region of the stellar bar. Figure\,\ref{fig:mass_accum}  provides such evolution inside an arbitrary chosen region, $|x| < 0.5$\,kpc along the bar, $|y| < 2$\,kpc along the bar minor axis, and within the layer of $z =\pm 3$\,kpc --- the disk thickness. The stellar bar growth period is associated with the increase of the central mass, in agreement with \citet{dubi09}. Hence, buckling is associated with an abrupt increase of the central mass by a factor of two, the subsequent oscillations, and continuing increase in the central mass thereafter, but at a much slower rate.

   \begin{figure*}
\center 
	\includegraphics[width=0.9\textwidth]{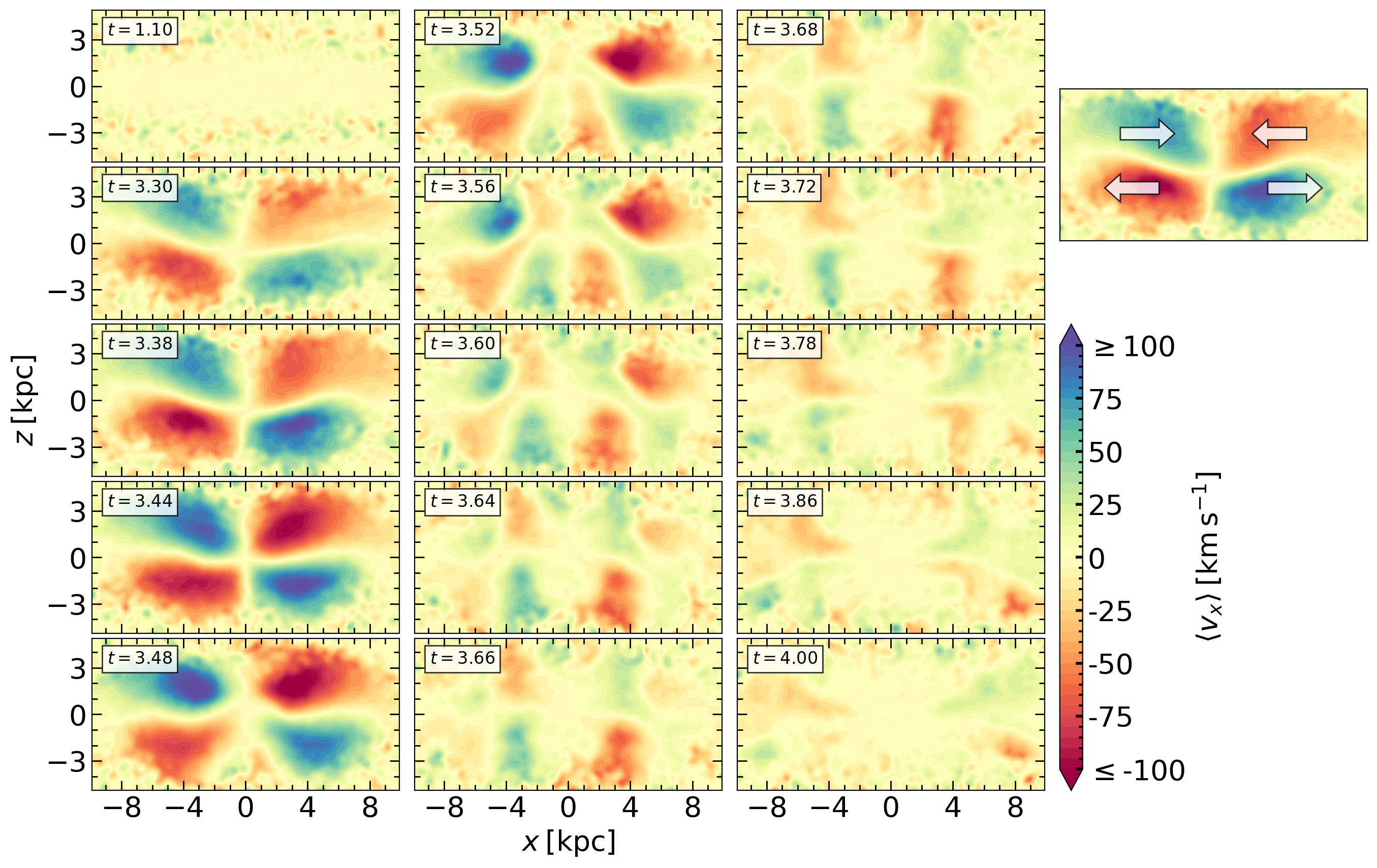}
    \caption{Evolution of the horizontal velocity along the bar within $|x| < 10$\,kpc, $|y| < 3$\,kpc, and $|z| < 5$\,kpc. Note that $x$- and $y$-axes corotate with the bar. The palette gives velocities in km\,s$^{-1}$. The right-hand insert displays the prevailing velocity trend: the buckling bell-shaped mode is pointing down, the compression prevails above the bell, and stretching prevails below the bell.  }
    \label{fig:vx_accum}
    \end{figure*}
  
    \begin{figure*}
\center 
	\includegraphics[width=0.9\textwidth]{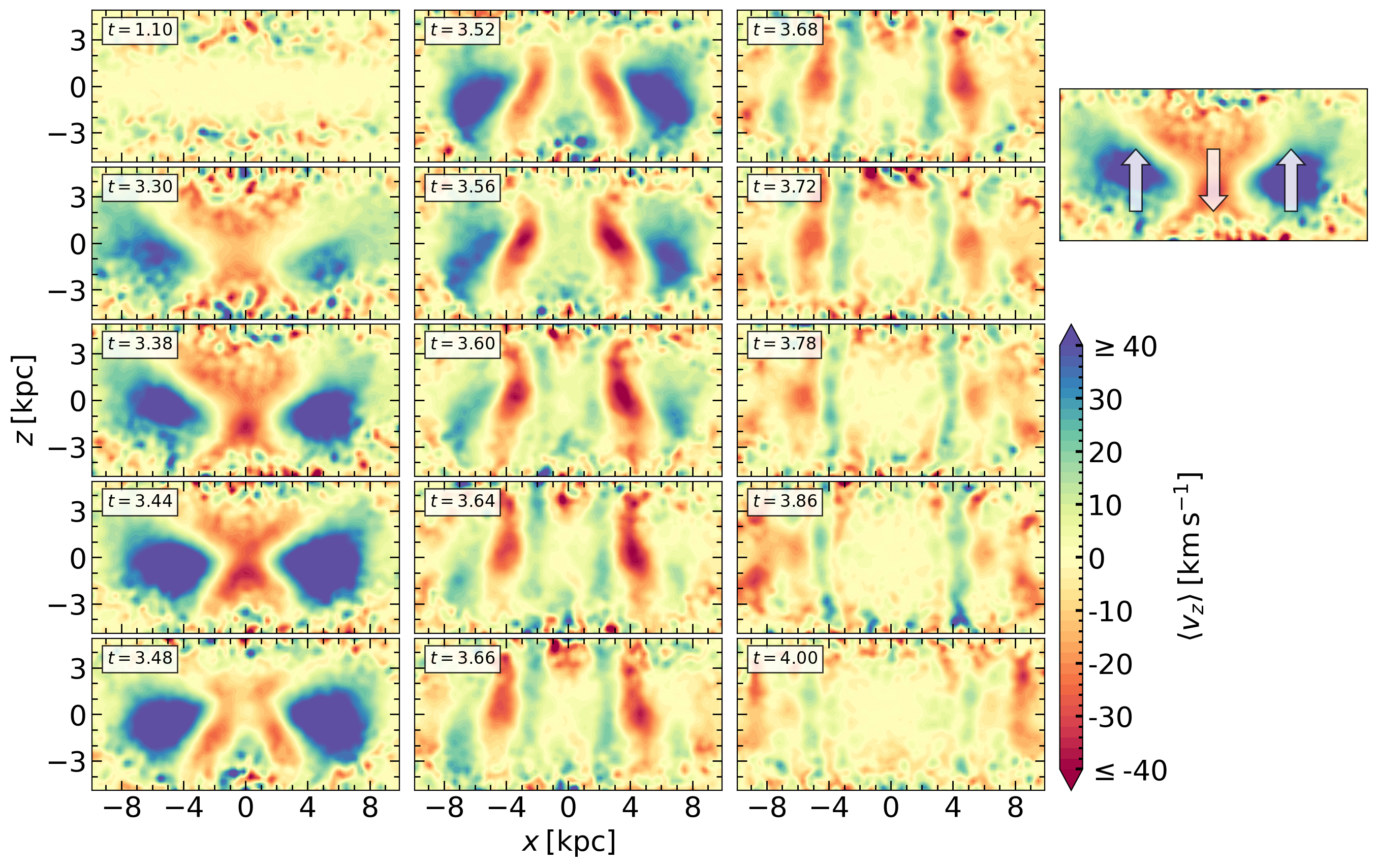}
    \caption{Evolution of velocity along the $z$-axis within the bar, $|x| < 10$\,kpc, $|y| < 3$\,kpc, and $|z| < 5$\,kpc. Note that $x$- and $y$-axes corotate with the major and minor axes in the disk plane respectively. The palette gives velocities in km\,s$^{-1}$. The right-hand insert displays the prevailing velocity trend: the buckling bell-shaped mode is pointing down. Note the similar appearance of circulation cells (e.g., Figure\,\ref{fig:vx_accum}), which expand along the bar after the buckling. }
    \label{fig:vz_accum}
    \end{figure*}  

    \begin{figure}
\center 
	\includegraphics[width=0.46\textwidth]{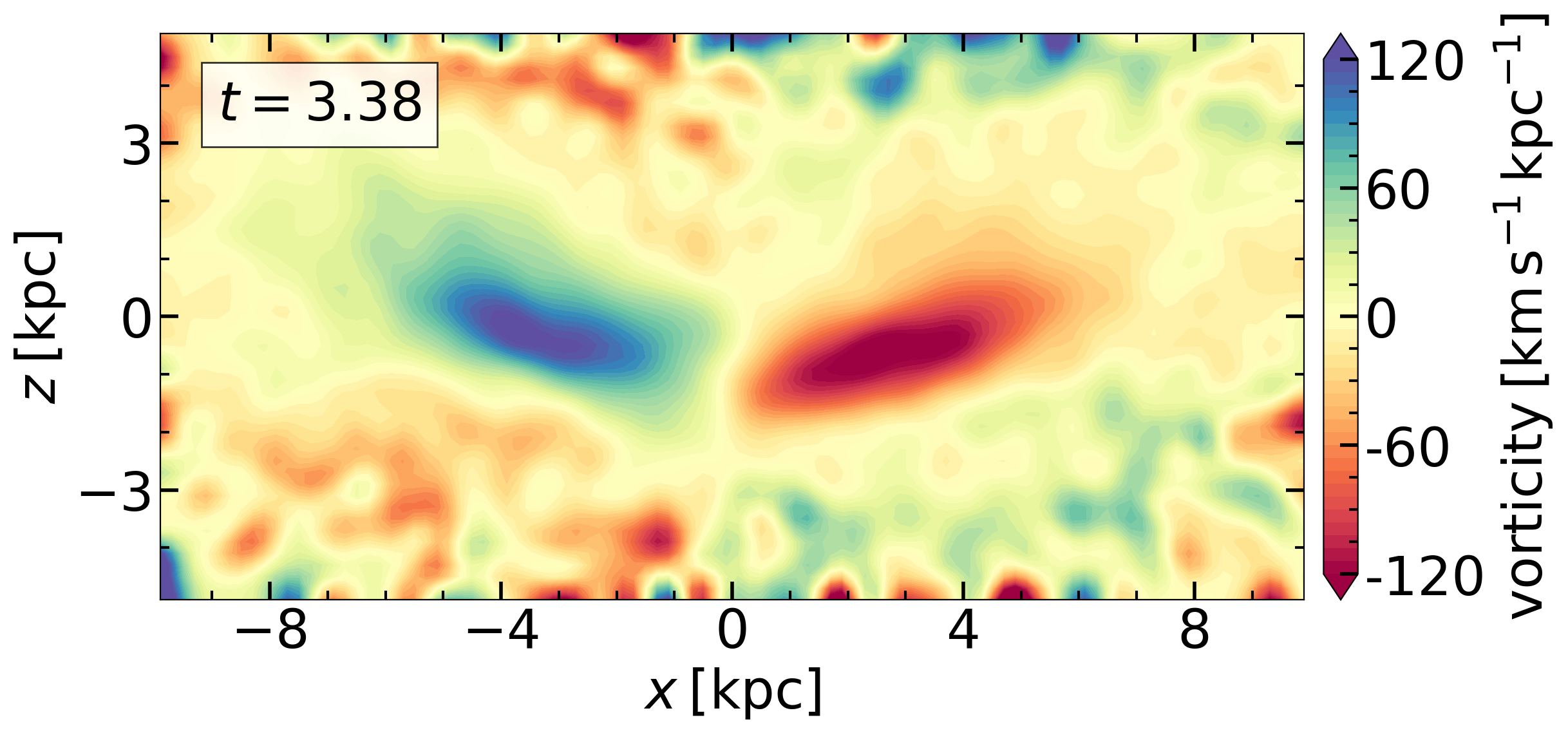}
    \caption{Snapshot of the projected vorticity component along the $y$-axis, ${\vec w}_{y}$, of the stellar bar calculated at $t=3.38$\,Gyr, to be compared with the circulation shown in Figs.\,\ref{fig:vx_accum} and \ref{fig:vz_accum}. The bar is horizontal and edge-on. Note that $x$- and $y$-axes corotate with the bar. The blue color corresponds to positive  vorticity along the $y$-axis in the $xz$-plane, and red color, to negative vorticity. See text for more details. 
    }
    \label{fig:vorticity}
    \end{figure}  

\subsection{Net velocity field in the stellar bar}
\label{sec:velocity}

To obtain further details of the central mass accumulation in the bar during buckling in Figure\,\ref{fig:mass_accum}, we analyze the velocity field along the bar (i.e., the $x$-axis which corotates with the major axis of the bar) as shown in Figure\,\ref{fig:vx_accum}, both in positive and negative directions, by displaying the representative snapshots, starting with $t=1.11$\,Gyr, which exhibits the uniform color around $v_{x} = 0$. This means that two opposite streams along the bar have negligible net velocity, with some noise present, and there is no net mass motion along the bar. As we aim at understanding the velocity field during buckling, the next snapshot at $t=3.0$\,Gyr exhibits a more complex field, as the bar midplane buckles down, i.e., we have a bell-shaped mode pointing down. The velocity field shows a preferred direction, $+x$ or $-x$. Below the bell-shaped distortion, velocity field represents stretching, while above it represents compression. 

The net velocity along the bar achieves its maximal value in the time period of $t\sim 3.3 - 3.48$\,Gyr, which reaches $\sim 100\,{\rm km\,s^{-1}}$ along $+x$ and $-x$ each. We note that compression velocities exceed those of stretching. This explains the mass growth in the center. We call a \textit{cell} this combination of compression and stretching velocity field.  Additional details of the velocity field become evident when following the evolution from snapshots at $t=3.48$\,Gyr till $3.86$\,Gyr. First, the single cell that existed prior to $t=3.48$\,Gyr bifurcates and at $t=3.48$\,Gyr we can already observe two such cells --- the outer ones which are strong and the inner ones which are weaker. Note that the inner cell net velocity is exactly opposite that of the outer cell. By $t=3.52$\,Gyr, the bell-shaped feature at the midplane flips over, the net flow becomes more complicated, and the cells interact diagonally.

By $t=3.64$\,Gyr, the outer cell dissolves, while the inner cell starts to expand outwards. We have measured this expansion of the cell between $t=3.66$\,Gyr and 3.78\,Gyr, and find it to be around $v\sim 30\,{\rm km\,s^{-1}}$. By $t\sim 4$\,Gyr, the cell dissolves completely, and the net mass flow along the bar vanishes. 

To get the full picture of circulation within the stellar bar, we measure the net mass flow along the $z$ coordinate. Figure\,\ref{fig:vz_accum} displays this motion by showing the average vertical velocity in each pixel. Again, well before the buckling we observe only noise, e.g., at $t\sim 1.1$\,Gyr. At $t=3.3$\,Gyr, the central region, during the downwards buckling,  displays a net velocity downwards, along $-z$ axis, $v_{z}\gtorder 10\,{\rm km\,s^{-1}}$. It is sandwiched by the upwards net current, $v_{z}\gtorder 30\,{\rm km\,s^{-1}}$. 

This motion along the $z$-axis can be unified with the previously discussed motion along the $x$-axis. We detect two circulation cells: the left one is clockwise, and the right one is counter-clockwise. Between $t\sim 3.44$\,Gyr and 3.48\,Gyr, we observe formation of additional circulation cell near the rotation axis, $x=0$, i.e., formation of the net flow along $+z$. By $t\sim 3.56$\,Gyr, this fully formed circulation starts to expand along the $-x$ and $+x$ axes. The sideways expansion proceeds with the same velocity of $\sim 30\,{\rm km\,s^{-1}}$, as in Figure\,\ref{fig:vx_accum}. Stellar orbits that can give rise to such circulation have been discussed in \citet[][see their Figure\,8]{heller96}, and are further presented in Appendix\,\ref{sec:append}.

These circulation cells have been never discussed in the literature and show that buckling instability is associated with increase in the mass concentration of stellar bar and with developing circulation patterns. These processes are the direct result of the action of self-gravity in the stellar `fluid,' and are absent in the classical description of the firehose instability in fluids and in plasma.
 
As an alternative description of the circulation in the bar at time of buckling, we use the vorticity,  $\vec w = \nabla\times (v_{ x}, v_{ y}, v_{ z})$. We are interested in the circulation in the $xz$-plane, hence we plotted its ${\vec w}_{ y}$ component in Figure\,\ref{fig:vorticity}, namely,

\begin{equation}
{\vec w}_{y} = \bigg(\frac{\partial v_{x}}{\partial z} - \frac{\partial v_{z}}{\partial x}\bigg) \hat y 
\end{equation}
The colors in Fig.\,\ref{fig:vorticity} correspond to the clockwise circulation on the left (blue color) and anti-clockwise circulation on the right (red color). 

For completeness, we have also calculated the divergence of the velocity field at the same time, and its agrees with the circulation shown in Figures\,\ref{fig:vx_accum} and \ref{fig:vz_accum}.
 
\subsection{Observational corollaries}
\label{sec:observe}

\begin{figure}
\center 
	\includegraphics[width=0.48\textwidth]{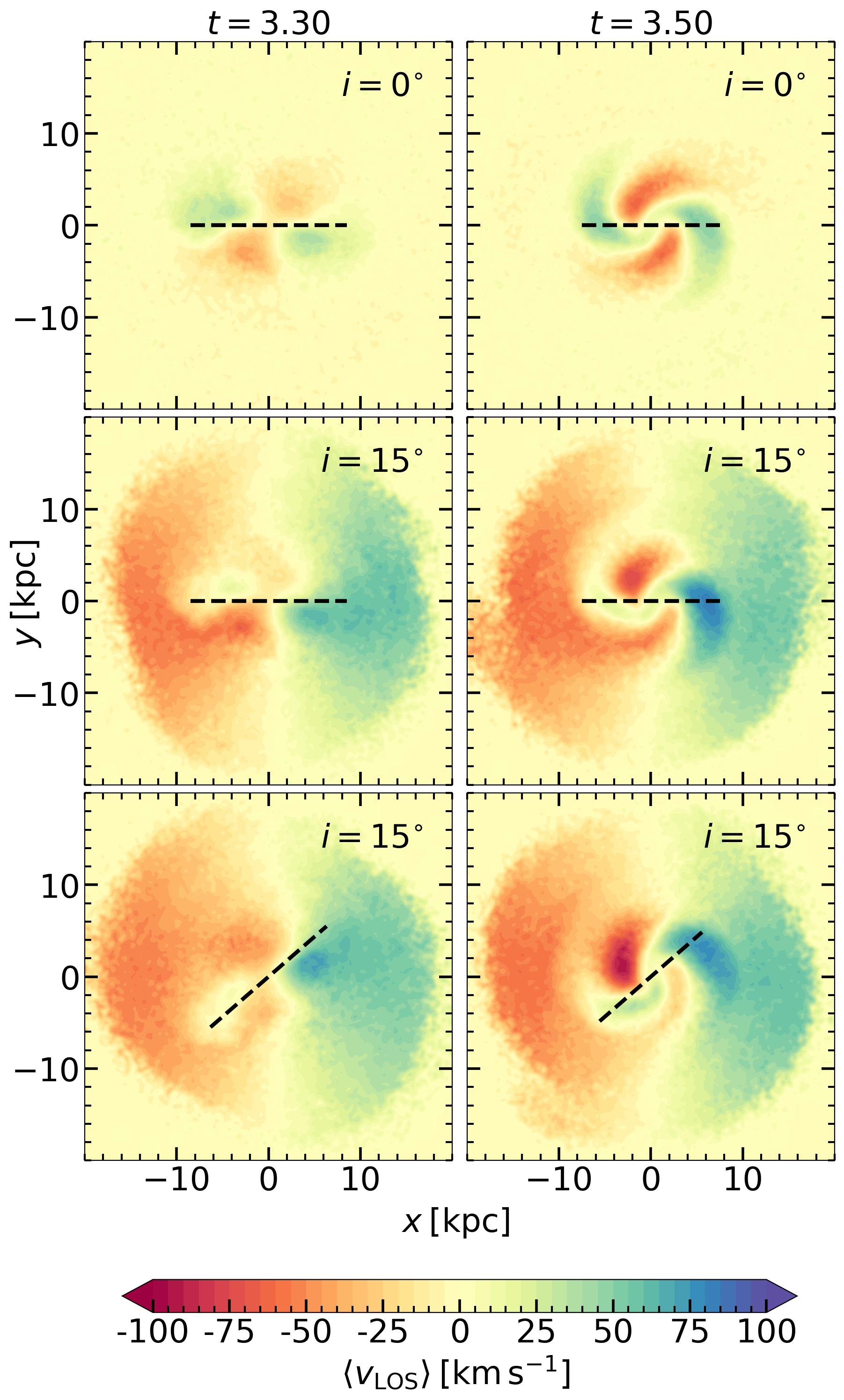}
	 \caption{The line-of-sight velocity field. The bar major axis and its extension are shown with a dashed line. The disk rotates counter-clockwise. The color palette shows the line-of-sight velocity. Top: for the face-on disk during maximal bending at $t=3.3$\,Gyr (left), and at $t=3.5$\,Gyr, with evolving circulation cells (right). Middle: line-of-sight velocity of a disk inclined by $15^\circ$ with respect to the bar major axis (x-axis). The top disk side is inclined forward. Note that the bar at $t=3.5$\,Gyr is not only bent along the $z$-axis but is deformed in the $xy$-plane as well --- this is typical during buckling. Bottom: same inclination of the disk as in the middle row, but with the bar inclined to $42^\circ$ to the $x$-axis.}
    \label{fig:vz_buck}
    \end{figure}

To analyze prospective observations of kinematics and morphology of stellar bars during the buckling process, we have calculated the velocity field along the rotation axis of the face-on and inclined disks, focusing on the maximal buckling amplitude at $t\sim 3.3$\,Gyr in our model. Figure\,\ref{fig:vz_buck} displays the line-of-sight velocity field of the face-on and inclined by $15^\circ$ disk, with the bar being horizontal, during the maximal buckling at $t = 3.3$\,Gyr and slightly thereafter, at $t=3.5$\,Gyr, when the circulation cells start to evolve, as discussed in section\,\ref{sec:velocity}. This signature of the buckling is in agreement with previous works \citep{lokas19a,lokas19b,xiang21}.

At small inclinations, $< 45^\circ$, the distortion in the velocity field is still detectable at the peak of the buckling. At $\gtorder 45^\circ$, the velocity footprint of the distortion is very weak, but appears to strengthen when the circulation cells expand radially. The velocity field affected by the buckling extends over the entire bar length. Note, that the bar has shortened during the buckling substantially, and the dashed line in Figure\,\ref{fig:vz_buck} corresponds to its instantaneous size. 
 
\subsection{The nonlinear phase of buckling instability}
\label{sec:buck}
 
     \begin{figure}
\center 
	\includegraphics[width=0.48\textwidth]{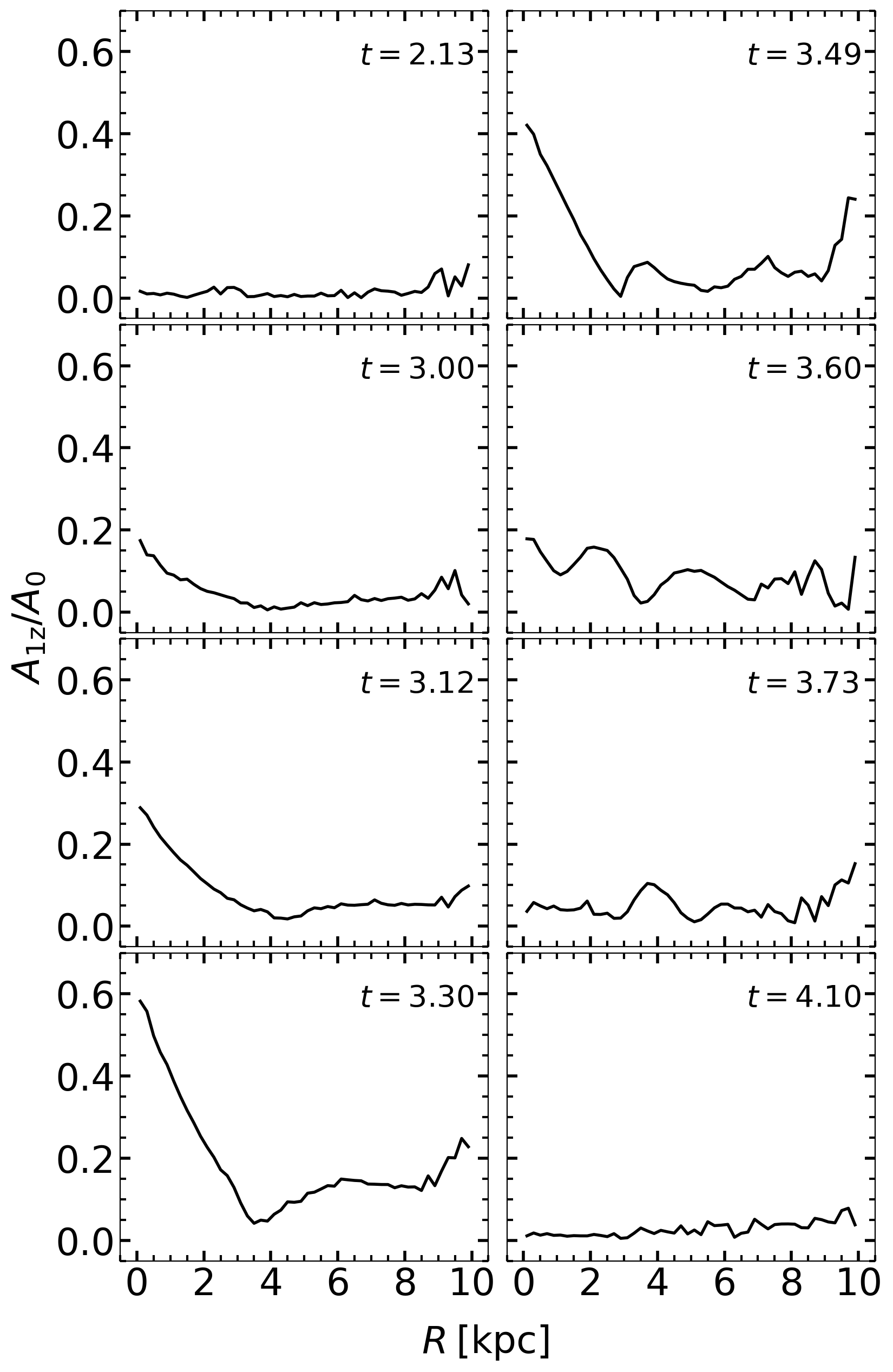}
    \caption{Evolution of the buckling amplitude radial profile, $A_{1z}(R)$, within the bar, $|x| < 10$\,kpc, $|y| < 3$\,kpc, and $|z| < 5$\,kpc.}
    \label{fig:a1zR}
    \end{figure} 
 
      \begin{figure}
\center 
	\includegraphics[width=0.48\textwidth]{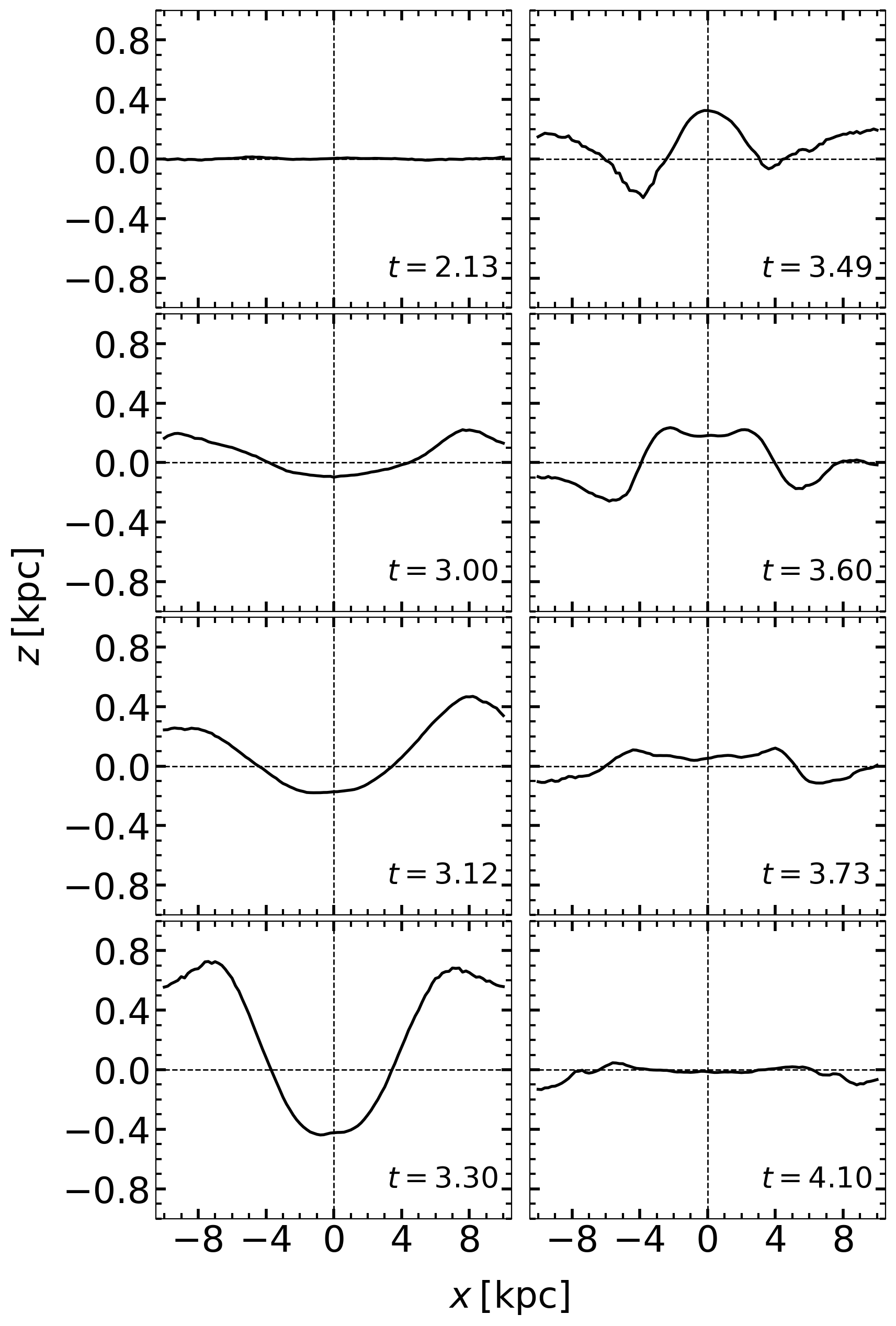}
    \caption{Evolution of the buckling amplitude of the total gravitational potential with the bar situated perpendicular to the line of sight. The potential minimum, i.e., of the warped Laplace plane, along the bar is calculated for $|x| < 10$\,kpc, $|y| < 3$\,kpc, and $|z| < 3$\,kpc region. The Laplace plane is determined when vertical gravitational acceleration is equal to zero. The origin point has been determined by the position of the center of mass of the whole system.}
    \label{fig:potentialMID}
    \end{figure}

We take a further look at the details of the buckling process. First, we look at at evolution of the buckling amplitude, $A_{1z}$, but focus on its radial distribution, compared to the radially integrated shown in Fig.\,\ref{fig:a2etc}. Plotting the radial properties of $A_{1z}(R)$ in Fig.\,\ref{fig:a1zR}, we observe that buckling only affects the region $r\ltorder 8$\,kpc, which defines 0.5 of the unstable wavelength, $\uplambda_{\rm buc}$ of the instability, and that its amplitude is quickly becoming nonlinear. We also observe that as the buckling amplitude declines, for $t > 3.5$\,Gyr, and the characteristic wavelength is halved --- exactly as we see in the radial and vertical net velocity diagrams (Figs.\,\ref{fig:vx_accum} and \ref{fig:vz_accum}). 
   
We define the Laplace surface (or a warped plane) as a surface where the potential has a minimum in the $z$-coordinate, or the surface where the $z$-component of acceleration vanishes \citep{dekel83}. The degree of nonlinearity of the buckling instability can also be measured using the curvature of the deformation of the Laplace plane in the system. This is shown in Fig.\,\ref{fig:potentialMID}, where the maximal distortion happens at $t\sim 3.3$\,Gyr. We have measured the slope of this distortion and found it making the angle $\theta\sim 9^\circ$ with the $xy$-plane at this time. Note, that the warped Laplace plane confirms $\uplambda_{\rm buc}\sim 16$\,kpc.

However, the gravitational potential is typically a slow varying function. We, therefore, recalculated the slope by fitting the projected density isocontours during the same time, and found to be $\theta\sim 19^\circ$. This value of $\theta$ clearly cannot be considered as a small one which does not affect the motion of stars along the bar in the central region. Which is also confirmed by a complicated net velocity fields during buckling in Figs.\,\ref{fig:vx_accum} and \ref{fig:vz_accum}.

The unstable wavelength of the buckling instability in our models is $\uplambda_{\rm buc} \sim 16$\,kpc. It can be obtained either from the bending of the Laplace plane in the gravitational potential or using the edge-on density contours of the bar. This should be compared to the disk (or bar) thickness at $R = 0$, which is $h\sim 6$\,kpc. So, in our case $\uplambda > h$. On the other hand, this wavelength is comparable to the bar or disk size. Hence the buckling instability, whatever is its cause is a global instability.

Finally, the mass influx towards the central region of the bar, at $t\sim 3.5$\,Gyr, basically doubles the mass there (e.g., Fig.\,\ref{fig:mass_accum}). This halves the buckling wavelength by about a factor of 2, as seen in Figs.\,\ref{fig:vx_accum}, \ref{fig:vz_accum} and \ref{fig:potentialMID}.  
 
\subsection{Resonant coupling}
\label{sec:resonance}
 
     \begin{figure}
\center 
	\includegraphics[width=0.46\textwidth]{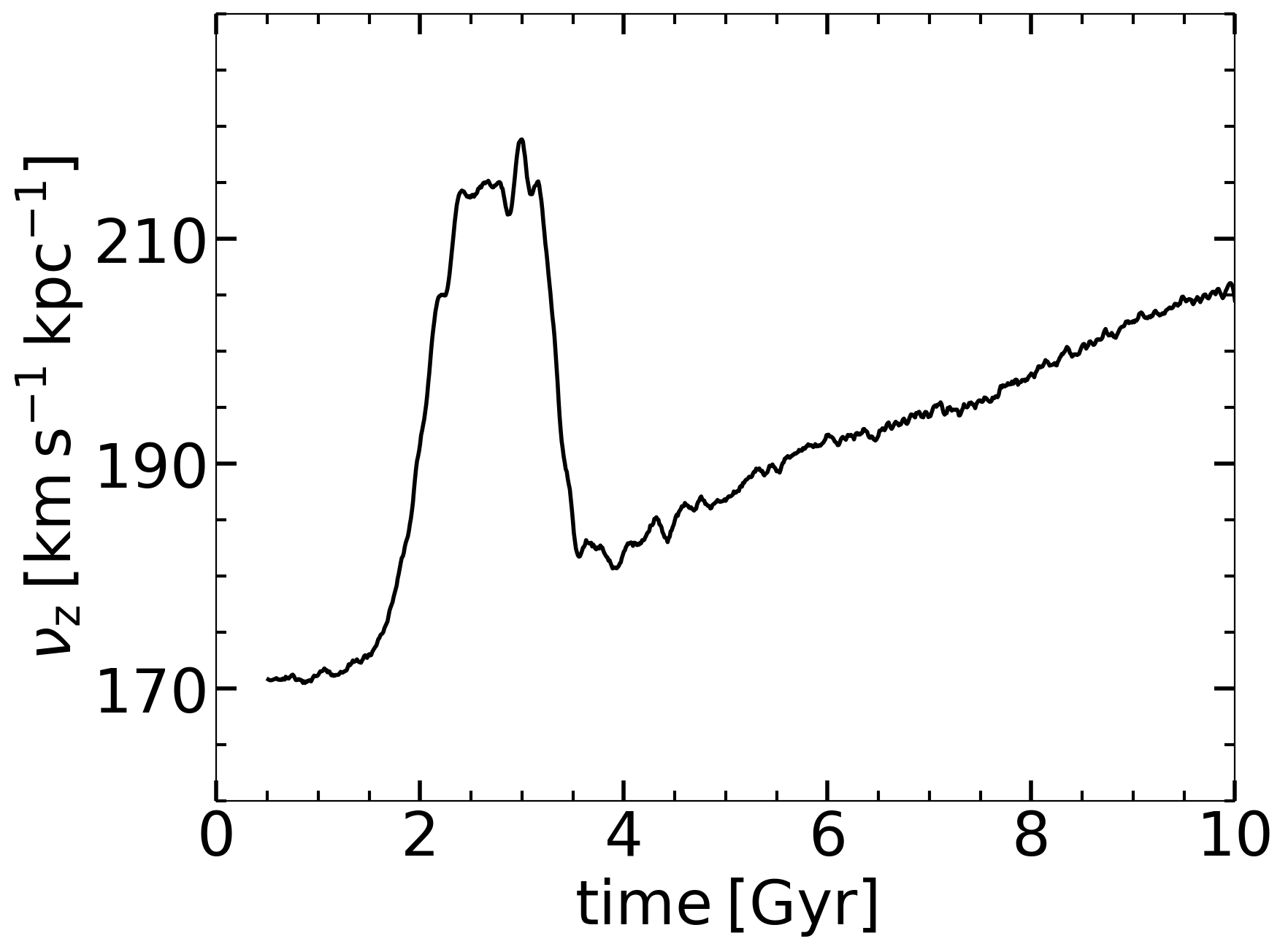}
    \caption{Evolution of the vertical frequency $\nu_{z}$ in central cylinder of 0.5\,kpc radius and $|z| < 3$\,kpc region for the standard model. It is calculated in the Laplace plane.}
    \label{fig:nu_center}
    \end{figure} 
 
\begin{figure}
\center 
	\includegraphics[width=0.48\textwidth]{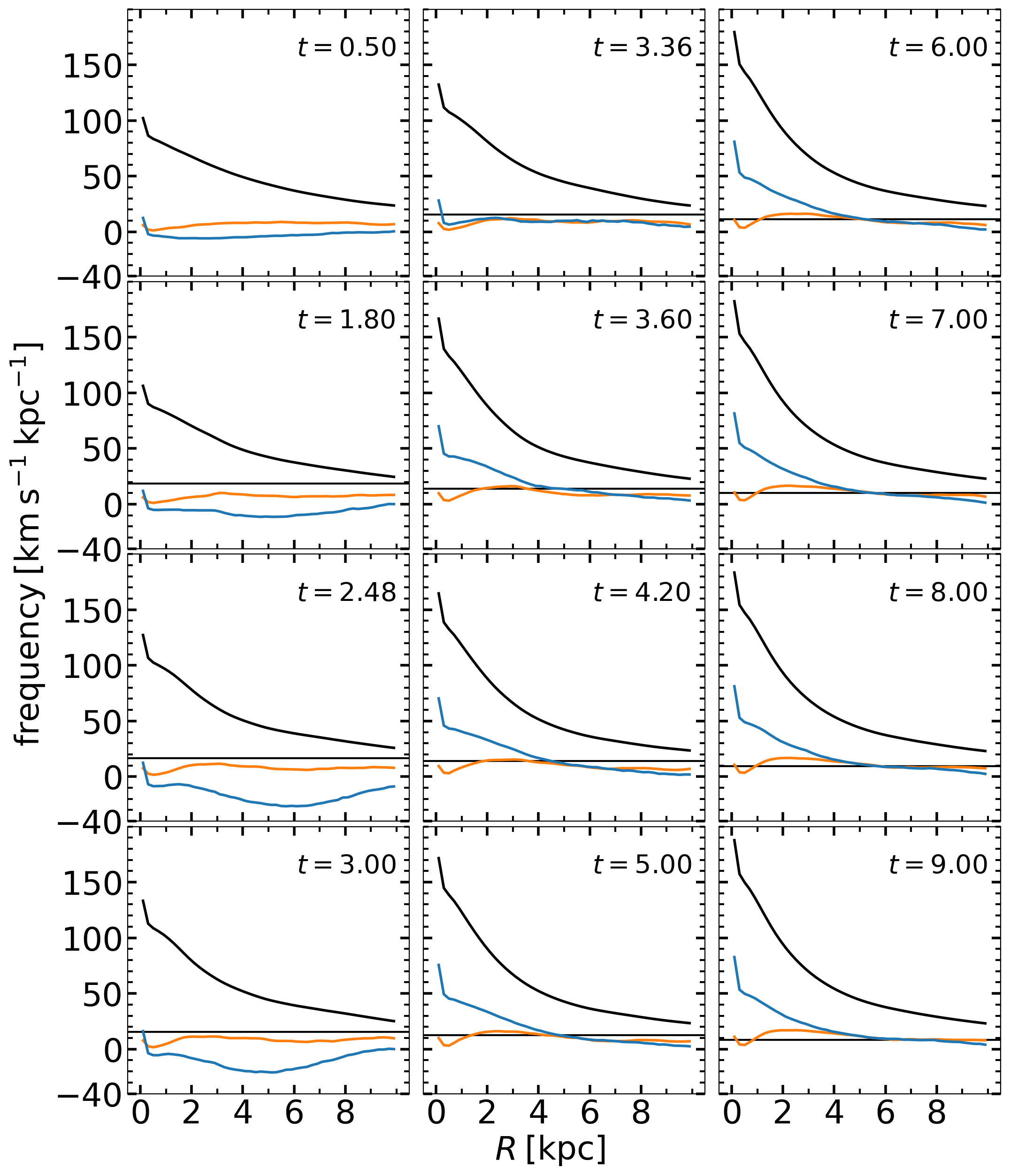}
    \caption{Evolution of the planar, $\Omega-\kappa/2$, and vertical, $\Omega-\nu_{z}/2$, ILRs. The black, orange and blue curves are the circular frequency $\Omega$, $\Omega-\kappa/2$, and $\Omega-\nu_{z}/2$, respectively. The frequencies $\Omega$, $\kappa$, and $\nu_{z}$ have been calculated within the bar $|x| < 10$\,kpc, $|y| < 3$\,kpc, and $|z| < 3$\,kpc. The horizontal line is the bar pattern speed, $\Omega_{\rm bar}$, at each time. The vertical frequency, $\nu_{z}$, has been calculated based on the vertical gravitational acceleration (see the text). Note, that the vertical ILR appears only with the vertical buckling, after $t\sim 3.0$\,Gyr, while the planar ILR appears after buckling, $t\gtorder 3.5$\,Gyr. The vertical frequency is calculated at the Laplace surface.}
    \label{fig:vert_res}
    \end{figure} 
 
So far we have focused on the firehose instability in the self-gravitating `fluid' as the physical model of vertical buckling of stellar bars. In the next step, we analyze the role of resonances in this buckling. We start with calculating evolution of the vertical frequency of stars in the bar, $\nu_{z}$. This is performed by calculating the position of the Laplace surface and its shape in the bar. Then calculating the vertical gradient of gravitational acceleration, yielding ${\rm d}a_{z}/{\rm d}z\sim -\nu_{z}^2$. Figure\,\ref{fig:nu_center} displays the evolution of this frequency in the central cylinder of radius $r=0.5$\,kpc and vertical extension of $|z| = 3$\,kpc. We observe  the growing $\nu_{z}$ in tandem with the growing bar. Then a short plateau and a sharp decline during buckling, with the subsequent growth, again in tandem with the bar.

While growing stages of vertical frequency can be easily understood as response to the increasing surface density in the bar (Fig.\,\ref{fig:mass_accum}), its sharp decline at the buckling requires some additional explanation. The stellar surface density of the bar does not decline at the buckling, but increases rather sharply. So why does $\nu_{z}$ declines? 

We have measured the slope of the vertical acceleration at the Laplace plane at $R=0$, and found that it flattens at end of the buckling. This leads directly to decrease in $\nu_{z}$. The reason for this flattening is that the curved bar plane at the end of buckling deflates (i.e., is washed out), decreasing the vertical gravity gradient at the center.  
 
In order to assess the importance of resonances, especially of the vertical ILR (vILR) on the buckling process, we have calculated the evolution of the planar and vertical profiles of $\Omega - \kappa/2$ and $\Omega - \nu_{z}/2$, respectively (Fig.\,\ref{fig:vert_res}). While these frequencies are strictly defined for the near-circular orbits, they still can serve as the first order approximation to actual orbital frequencies in strongly nonlinear stages of evolution. 

In our model, the vILR appears only at the beginning of the buckling in the Laplace plane, $t\sim 3.0$\,Gyr. Its position is changing with time --- first it can be recognized in the central 100\,pc. With the gravitational softening of 25\,pc only, this observation appears to be reliable. The vILR is slowly moving outwards and stagnates at $\sim 6$\,kpc --- which approximately corresponds to the extension of the boxy/peanut-shaped bulge. This observation is important because the original analysis of formation of such bulges has been performed at an epoch characterized by a much weaker computing power \citep{comb90,pfen91}, while the appearance of the vILR has been determined only after buckling at a single time.

The planar ILR in Fig.\,\ref{fig:vert_res} appears around $t\sim 3.5$\,Gyr, hence only after buckling. It shows up at $R\sim 2$\,kpc as a single planar ILR, and then separates into inner and outer ILR, i.e., IILR and OILR. The IILR moves inwards and ends at $R\sim 1$\,kpc, while OILR moves outwards and ends up at the position of the vILR. In fact, almost immediately after the OILR has been recognized, after $t\sim 4.0$\,Gyr, it moves out and coincides with the vILR. This is significant, as the two resonances, the planar and the vertical overlap and can work in tandem. Moreover, an extended region where the ratio $(\Omega - \kappa/2)/(\Omega - \nu_{z}/2)\sim 1$, is maintained after buckling.

We note that the resonances discussed in this section represent the {\it linear} resonances, the ILRs. These are defined for small perturbations of circular orbits. ILRs defined this way, if they exist, should lead to the appearance of the family of $x_2$ orbits, which are oriented perpendicular to the bar major axis \citep{cont80}. The fully nonlinear analysis which is performed in the next section and the resulting families of orbits which are displayed in Appendix\,\ref{sec:append}, did not detect such family of orbits as present in our model --- a well known occurrence in orbital dynamics. However, the nonlinear orbital analysis has indicated that both vertical and planar 2:1 resonances are present in the model, as shown in Section\,\ref{sec:versus}. To preclude any confusion, we use the ILR definition only when the $x_2$ family of stellar orbits is present, and, in the absence of this family, refer to the resonances simply as planar and vertical 2:1 resonances. In the next section, we define the nonlinear frequencies which are used for detecting the nonlinear resonances only.

 \section{Buckling: firehose instability versus resonance triggering}
\label{sec:versus}

To summarise our results so far --- we have analyzed arguments for two alternative triggers of the vertical buckling instability in stellar bars. In the next step, we confront these triggers and attempt to make conclusions on their relative importance for the discussed instability. Before we proceed, we describe the general conditions in the disk with growing bar.

In the previous section, we have defined the linear frequencies, $\kappa$ and $\nu_{ z}$. Here we define their non-linear extensions. We use $\Omega_{R}$  as an extension of $\kappa$, where we assume an arbitrary amplitude oscillations along the cylindrical coordinate $R$. The nonlinear frequency $\Omega_{x}$ has been defined along the bar major axis in the rotating $\Omega_{\rm bar}$ frame. We emphasize that $\Omega_{R}$ differs from $\Omega_{x}$. We also define the extension of $\nu_{z}$ being $\Omega_{z}$.  

Note that these definitions can result in 2:1 and other resonances, which will not bring in the $x_2$ family of orbits. Indeed, using the characteristic diagram (section\,\ref{sec:append}), we do not detect this family of stellar orbits. But we do detect resonances based, e.g., on $\Omega_{z}/\Omega_{x}$ and $\Omega_{R}/\Omega_{x}$ being rational numbers. In this case, we do not call them Lindblad resonances, but simply use, say, 2:1 ratios.

 \subsection{Coupled driven oscillators and stability}
\label{sec:oscillate}

In the galaxy center, the disk dominates the gravitational potential, and has an exponential or near-exponential density profile (e.g., Eq.\,\ref{eq:rho_d}). On the larger scale, the system potential is of course dominated by the DM halo, resulting in the flat rotation curve and approximated by the logarithmic potential. 
\begin{equation}
 \Phi = \frac{v_0^2}{2}\log\left(1 + \frac{x^2}{a^2} + \frac{y^2}{b^2} + \frac{z^2}{c^2} \right),  
 \label{potent}
\end{equation}
where $v_0$ is the asymptotic circular velocity and $a$, $b$, and $c$ the core potential semi-axes. 
Indeed to first order in $x^2$, $y^2$, and $z^2$ such a potential expands to a harmonic potential,
\begin{equation}
 \Phi_\textrm{core} = \frac{v_0^2}{2} \left(\frac{x^2}{a^2} + \frac{y^2}{b^2} + \frac{z^2}{c^2} \right) = 
   \frac{\Omega_{x}^2}{2} x^2 + \frac{\Omega_{y}^2}{2} y^2 + \frac{\Omega_{z}^2}{2} z^2 ,  
 \label{harmonic_potent}
\end{equation}
where $\Omega_{x} = v_0/a$, and so on.

As such it forms a central disk density core of size $R_0 \approx (a^2+b^2)^{1/2}$, i.e., a harmonic core with the size approximately that of the disk lengthscale. 

While typical cosmological halos are universally triaxial, as shown in numerical simulations \citep[e.g.,][]{allgo06}, the baryonic collapse results in the washing out this triaxiality, especially the DM halo prolateness \citep{bere06}. Having in mind this and that our potential is {\it oblate} due to the presence of the galactic disk, we fitted $\Phi$ in Eq.\,(\ref{potent}) at $t=0$. The resulting harmonic core radius was obtained as $R_0 \sim 2.7$\,kpc, i.e., only slightly modified with respect to the exponential stellar disk scalelength.

An {\it axisymmetric} oblate harmonic core, e.g., prior to the bar development, in 3-D is characterized by two independent orbital frequencies, as $\Omega_{x} = \Omega_{y}$, the planar and vertical frequencies, which are generally incommensurable, i.e., the ratio $\Omega_{z}/\Omega_{x}$ is an irrational number. This is an example of an anisotropic oscillator. 

The effective potential in the rotating frame takes into account the centrifugal force potential contribution $-\Omega_\mathrm{bar}^2 (x^2+y^2)/2$, where $\Omega_\mathrm{bar}$ is the bar pattern speed.  The effective core potential in the rotating frame becomes 
\begin{equation}
 \Phi_\textrm{bar} = 
 \frac{1}{2}\left(\Omega_{x}^2-\Omega_\mathrm{bar}^2\right) x^2 + \frac{1}{2}\left(\Omega_{y}^2-\Omega_\mathrm{bar}^2\right) y^2 + \frac{1}{2}\Omega_{z}^2 z^2 .  
 \label{rotating_harmonic_potent}
\end{equation}
This leads to three independent frequencies which are generally incommensurable \citep[e.g.,][]{elzant02}.

Following the theory of simple oscillators, when the natural frequency of the oscillator is smaller than the perturbing force frequency, it will respond out-of-phase with the force. If, however, the natural frequency of the oscillator is larger than the perturbing force frequency, the oscillator can respond in-phase with the force. For a single oscillator, this is as far as it goes. However, if we deal with a family of coupled oscillators subject to the external perturbing force, the situation is by far more complex. 

Suppose we have such a group of uncoupled physical oscillators, with a natural frequency $\omega_0$. Suppose they are driven with a harmonic force with an amplitude $f_0$, i.e., $f(t) = f_0 {\rm cos} (\omega_{\rm f}t)$, where $\omega_{\rm f}$ is the driving  frequency. If the typical mass and size of the oscillator are $m$ and $l$, and it is subject to a restoring gravitational acceleration $g$, one can define a dimensionless parameter $\gamma$,
\begin{equation}
\gamma = \frac{f_0/m l}{\omega_0^2} \equiv \bigg(\frac{\omega_{\rm f}}{\omega_0}\bigg)^2.    
\end{equation}
When $\gamma < 1$, the perturbation is relatively small, as $f_0 < mg$. The linear regime corresponds to $\gamma \ll 1$, and the motion is harmonic with the periodicity of the driving force. A single attractor exists in this case. For $\gamma \sim 1$ and slightly above it, additional harmonics appear, which decay after transients go away. For $\gamma$ still increasing,  the chaotic response will follow.

What follows is that for $\gamma \ltorder 1$, the system of these oscillators can respond in tandem, because a single harmonic prevails. If some additional `glue' exists, e.g., connecting springs, it will assure a cohesive response. If, however, $\gamma > 1$, this cohesiveness will be washed away, and no collective response will occur. Such cohesiveness can be supplied by gravitational interactions between the oscillators, as in the case of the stellar `fluid,' i.e., by the self-gravity, and we discuss it below.  

 \subsection{Stellar bars: coupled driven oscillators}
\label{sec:bar}

We define the actual nonlinear oscillations of stars along the $x$ and $z$ axes in the following way. Tracing stars along their orbits in the live potential, we determine the period of the $i$-orbit as the time between two successive apocenters of this orbit, $T_{i}$, and their `instantaneous' frequencies as $\Omega_{i,x} = 2\pi/T_{i,x}$ and $\Omega_{i,z} = 2\pi/T_{i,z}$. We omit the index $i$ in the text to simplify it.

If the radial oscillations of stellar particles serve as a driving force for vertical buckling, the frequency of this force is $2\Omega_{x}$, because the star passes twice during one full horizontal oscillation over the position of the vertical oscillating particle. Therefore, $\Omega_{z}$ corresponds to the natural frequency $\omega_0$ discussed in the previous section, and $2\Omega_{x}$ corresponds to $\omega_{\rm f}$. Moreover, the self-gravity of stars plays the role of the cohesive force.  

The main conclusion from this discussion is that when $\Omega_{ z}/2\Omega_{ x} > 1$, the system can respond in tandem, and the cohesiveness will assure  that the bar will buckle vertically. On the other hand, when $\Omega_{ z}/2\Omega_{ x} < 1$, the response will be chaotic, and each orbit will respond out of phase with the other, thus ensuring stability against buckling. In this latter case, the bar can still thicken and populate the BAN/ABAN orbits without buckling. 

\begin{figure*}
\center 
	\includegraphics[width=1.0\textwidth]{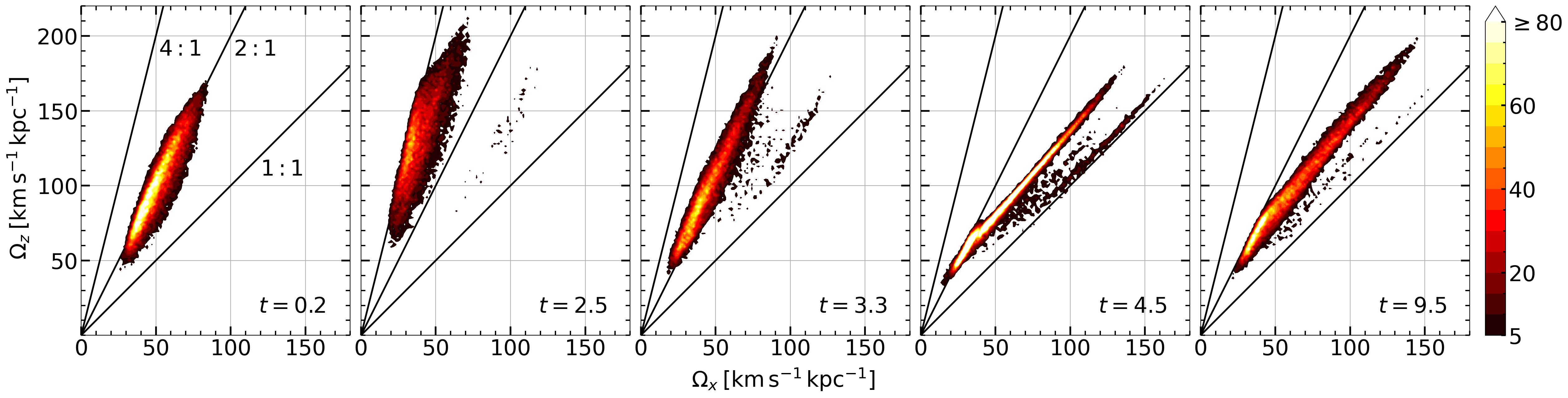}
	 \caption{Evolution of the resonant frequencies, $\Omega_{ z}-\Omega_{ x}$, in the stellar bar, between $t=0$ and $t=10$\,Gyr. The sample of 30,000 particles has been selected randomly from the bar volume at $t=2$\,Gyr, close to the maximum of the bar $A_2$ amplitude. Shown are the surface density contours (see the color palette for details) at times $t=0$, 2.5\,Gyr, 3.3\,Gyr, 4.5\,Gyr and 9.5\,Gyr. The diagonal lines represent (vertical) resonances in the $zx$-plane: 1:1 resonance (the lowest line), 2:1 resonance (the middle line), and 4:1 resonance (the top line). The associated animation displays this evolution between $t=2.2$\,Gyr and 4.5\,Gyr. This corresponds to the strong bar growth, its buckling and the end of the buckling. During the remaining time, the slope increases only negligibly from $\sim 1.15$ to 1.19 at 9.5\,Gyr. But as the bar resumes its secular growth, the particle distribution diffuses to higher $\Omega_{ z}$ and $\Omega_{ x}$ along the same slope.}
    \label{fig:sigmas2}
    \end{figure*}

We have randomly selected 30,000 stellar particles from the bar volume\footnote{The bar volume has been approximated by an ellipsoid with the three principal axes obtained from the bar size and its ellipticity, and assuming that the bar has the disk thickness before buckling.} at $t = 2$\,Gyr, close to the bar maximal amplitude before buckling, and calculated the evolution of their actual nonlinear frequencies, $\Omega_{z}$ and $\Omega_{x}$, in the following way: (1) To facilitate the determination of the orbital apocenters of particles, we increase the frequency of dumping of particle positions to every 1 Myr, and (2) by interpolating between the particle positions using a cubic spline algorithm. This procedure results in smooth orbits and instantaneous  $\Omega_{z}$ and $\Omega_{x}$ frequencies. Lastly, we plot the positions of these particles in the frequency space in Fig.\,\ref{fig:sigmas2}.  

We have modeled the evolution of the above sample of particles in the $\Omega_{ z}-\Omega_{ x}$ plane between $t=0$ and $t=10$\,Gyr. A few snapshots of this evolution are presented in Figure\,\ref{fig:sigmas2}. We also present the animation of the particle evolution between $t=2$\,Gyr and 4\,Gyr, covering the entire phase of buckling. On this plane, particles with small $\Omega_{ x}$ lie at large $R$, at the outer edge of the bar, while particles with progressively larger $\Omega_{ x}$, lie at smaller $R$, respectively. The vertical resonances, 1:1, 2:1, and 4:1 (i.e., the vertical Ultra-Harmonic resonance) are shown with diagonal lines. In terms of the Jacobi energy, $E_{ J}$, particles within a narrow range of Jacobi energy form a vertical strip, with more negative $E_{ J}$ situated at larger $\Omega_{x}$ and smaller $R$.  

At $t=0$, the particles are distributed along the vertical 2:1 resonance line, with the peak density on this line. This, however, is not obligatory but a simple coincidence. For disks which have larger vertical dispersion velocities initially, the particles lie just below the vertical 2:1 resonance line and move up with the bar growth across the 2:1 line. While for disk which are colder, more particles are situated above the vertical 2:1 resonance initially. This trend affects the timing of the buckling --- colder bars buckle earlier than the hotter ones.

At $t=2.5$\,Gyr (i.e., at the maximum of $A_2$), the particles distribution has moved up, and is situated between the vertical 4:1 and 2:1 resonance lines. At $t = 3.3$\,Gyr (i.e., the maximal buckling amplitude), the particles lie on a narrow strip just touching the vertical 2:1 resonance line and following its slope with the $x$-axis. At $t=10$\,Gyr, the slope of the particle distribution corresponds to $\Omega_{ z}/2\Omega_{ x}\sim 1.19$, below the vertical 2:1 resonance line, which is slightly increased from the slope $\Omega_{ z}/2\Omega_{ x}\sim 1.15$ at $t=4.5$\,Gyr.  Hence, the bar becomes unstable after reaching a particular strength, until the buckling time, $t\sim 3$\,Gyr.  

Our results indicate that the particles cross $\Omega_{ z}/2\Omega_{ x} = 1$  line starting at $t\sim 3.3$\,Gyr, which corresponds to the maximal buckling. This line corresponds to the specific commensurability region of the vertical 2:1 resonance, which provides the resonant coupling between these two frequencies.   At this moment particles touch the 2:1 resonance line. The larger $\Omega_{ x}$ particles, i.e., most negative $E_{ J}$, cross this line first, and the rest of them within $\sim 0.5$\,Gyr. This time interval corresponds to the buckling described by various methods in our Figs.\,\ref{fig:mass_accum} -- \ref{fig:potentialMID}. It is also in a good agreement with the evolution of the vertical frequency $\nu_{ z}$ in the central region of the bar of Fig.\,\ref{fig:nu_center}.

After the bar has reached its maximal strength, $\Omega_{ x}$ is steadily increasing while $\Omega_{ z}$ is steadily decreasing. The ratio $\Omega_{ z}/\Omega_{ x}$ is, therefore, steadily decreasing as well. Note, that after $\sim 4.5$\,Gyr, the slope of the particle distribution is nearly constant, and increases only negligibly from $\sim 1.15$ to $\sim 1.19$ at $t=10$\,Gyr. As the bar resumes its growth after buckling, the particle distribution diffuses slowly along the same diagonal and also `puffs' up in the perpendicular direction, without changing its slope in the $\Omega_{ z}-\Omega_{ x}$ plane. 
 
\begin{figure}
\center 
	\includegraphics[width=0.48\textwidth]{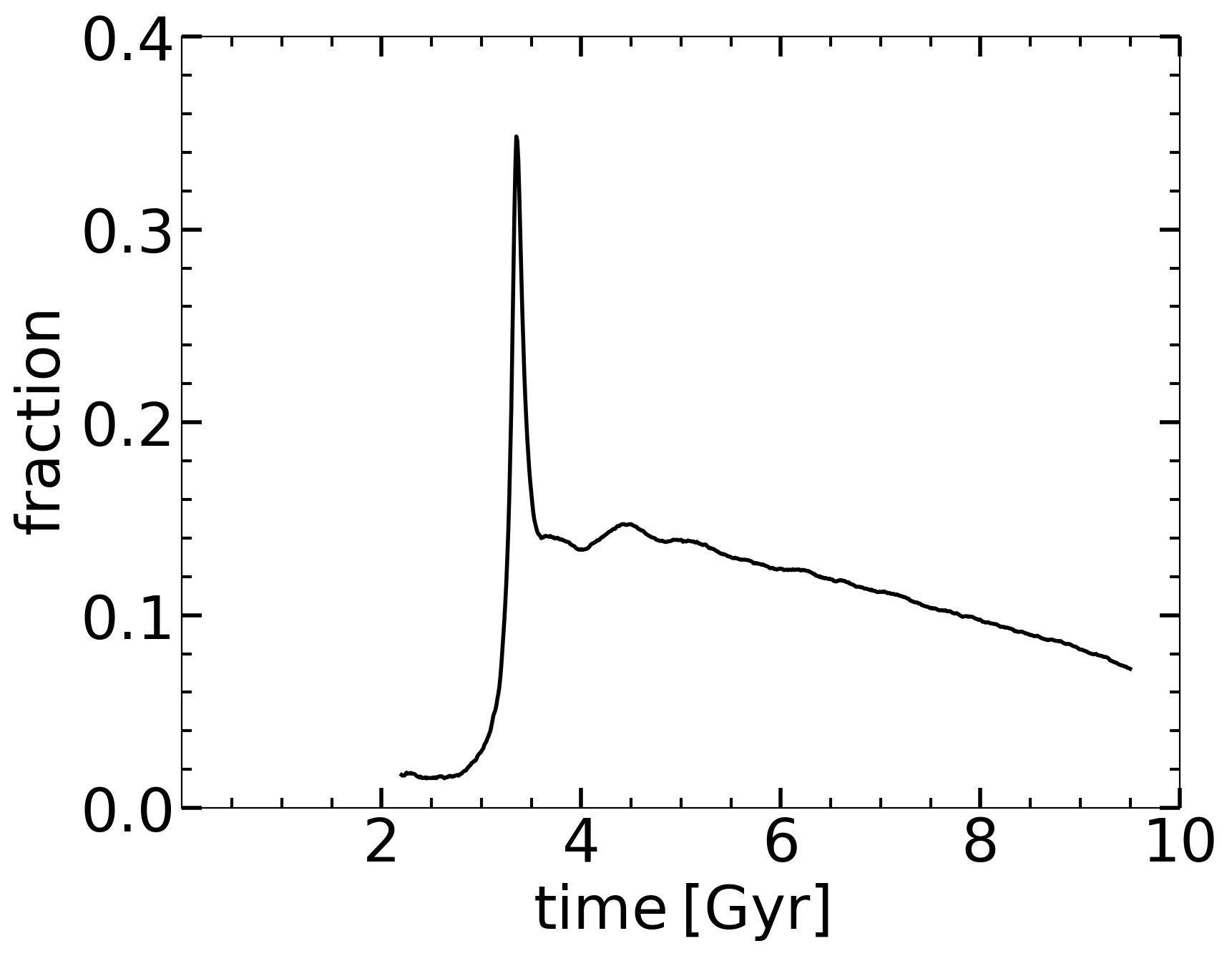}
	 \caption{Evolution of the fraction of the bar particles that are swept by the vertical 2:1 resonance as a function of time, for $t\ge 2$\,Gyr (see also Fig.\,\ref{fig:sigmas2} and the animation of this Figure). The width of the vILR is taken assuming that $\Delta\Omega_{z} = \Delta\Omega_{x} = \pm 5\,{\rm km\,s^{-1}\,kpc^{-1}}$.}
    \label{fig:cross}
    \end{figure} 

To determine the trapping of particles by the vertical 2:1 resonance, as they are swept by this resonance, we plot the evolution of fraction of particles in the stellar bar within this resonance width as a function of time. Figure\,\ref{fig:cross} displays this fraction assuming the width of the resonance as $\pm 5\,{\rm km\,s^{-1}\,kpc^{-1}}$. We  observe a strong peak at $t\sim 3.3$\,Gyr, exactly at the peak of the buckling amplitude shown in previous Figures. A substantial fraction of the bar particles, $\sim 1/3$ are associated with this peak and appears to be sufficient to assure the cohesiveness of the buckling, i.e., its collective response.

As Figure\,\ref{fig:cross} displays the evolution of the orbital fraction crossing the vertical 2:1 resonance, after majority of the particles cross the resonance width, some outer particles remain within the resonance (see Fig.\,\ref{fig:sigmas2}, particles with lower $\Omega_{ x}$), with their number slowly decreasing with time, displayed by a slowly decreasing curve.   

A fraction of particles trapped in the vertical 2:1 resonance is found also trapped by the {\it planar} 2:1 resonance. We define the $xy$-planar 2:1 resonance by $\Omega_{R}$ which is an extension of $\kappa$, i.e., when $\Omega_{ R}/2\Omega_{ x} = 1$. Note that $\Omega_{ x} = \Omega - \Omega_{\rm bar}$ in the bar frame. Simultaneous trapping by vertical and planar resonances is a clear signature of coupling between the motion in the $xy$- and $zx$-planes.

\begin{figure}
\center 
	\includegraphics[width=0.48\textwidth]{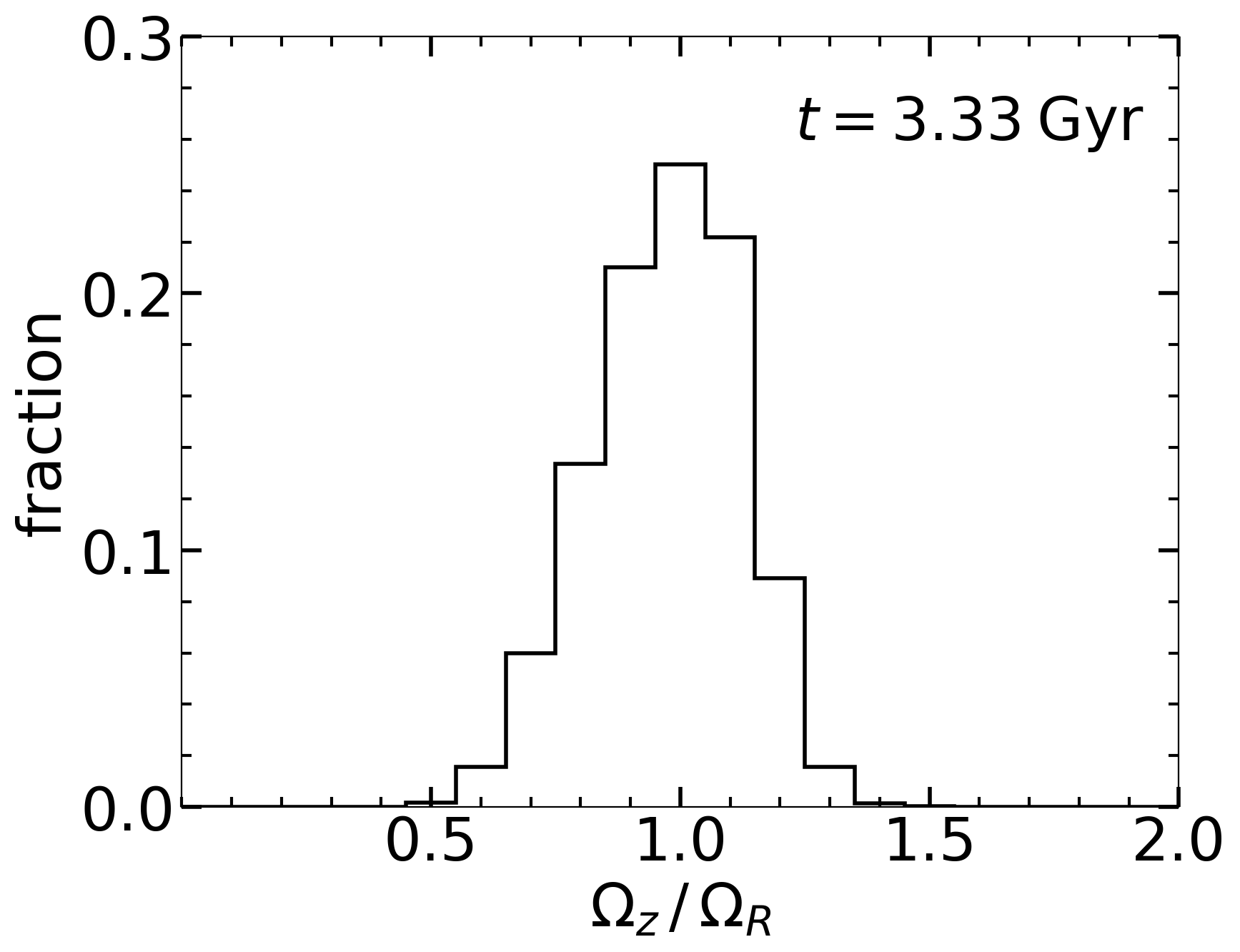}
	 \caption{Overlapping planar and vertical 2:1 resonances. Fraction of stellar orbits trapped by the planar 2:1 and vertical 2:1 resonances at the time of maximal buckling amplitude. More than a quarter of the particles trapped in the former resonance are trapped by the vertical resonance. }
    \label{fig:trapping}
    \end{figure}

We have analyzed the planar and vertical 2:1 resonances overlap. Figure\,\ref{fig:trapping} displays this overlap at $t=3.3$\,Gyr, when $\Omega_{ z}/\Omega_{ R} = 1$. The bin size is chosen to be the resonance width, $\Omega_{ R} = 2\Omega_{ x} \pm 5\,{\rm km\,s^{-1}\, kpc^{-1}}$. More than a quarter of the total number of particles selected at $t=2$\,Gyr are trapped by the overlapping planar and vertical 2:1 resonances. This can be sufficient to ensure the cohesiveness of orbital response and result in the buckling instability.  

A separate issue is that in all of our models, the particle distribution touches the vertical 4:1 resonance line in Fig.\,\ref{fig:sigmas2}, with negligible number of particles that cross it. This phenomenon can be analogous to the well known property of bars to extent to only the planar UHR, if the bar is very strong, and cross it towards the CR in other cases. This effect must be analyzed in a more depth.

\section{Discussion}
\label{sec:discussion}

We have analyzed models aiming at the cause of the vertical buckling instability in stellar bars by focusing on the nonresonant firehose instability and resonant vertical excitation of stellar orbits. So far, analytical methods have been applied exhaustively to the linear stage of this instability. They assumed that the velocity field in the bar is not affected by the buckling, as well as ignored additional evolutionary processes in the system. However, the nonlinear stage can only be investigated using numerical methods and we resort to this approach.

For this purpose, we used the collisionless $N$-body system of a stellar disk embedded in the DM halo. The stellar bar has developed in a few disk rotations, and we followed the mass redistribution in the growing bar and the associated velocity field there. Vertical buckling developed after the bar reached its maximal strength.

Our main results refer to conditions which favor instability which is driven by nonresonant and resonant effects, and are as following:

\begin{itemize}
    
\item The buckling is associated with the abrupt increase in the central mass concentration in the bar by about a factor of two.

\item Close to the buckling time, $\pm 0.3$\,Gyr, the velocity field shows a preferred direction towards the rotation axis, with the amplitude of $\sim 100\,{\rm km\,s^{-1}}$ in each direction. Below the bell-shaped distortion, the velocity field represents stretching, while above it represents compression, with the latter exceeding the former. This combination of compression and stretching forms a velocity cell.  

\item Immediately past the maximal buckling, the velocity cell bifurcates, forming inner and stronger outer cells. The velocities are reversed between the inner and outer cells. As they evolve, the cells interact diagonally. In a very short time, the outer cells dissolve, while the inner cells start to expand outwards with a typical velocity of $\sim 30\,{\rm km\,s^{-1}}$. By $\sim 0.5$\,Gyr after the maximal buckling the cells dissolve.

\item During buckling, we detect a complicated velocity field along the rotation axis as well. On the $z$-axis, the net vertical velocity is $\sim 10-15\,{\rm km\,s^{-1}}$ in the direction of the bell-shaped distortion. This motion, in tandem with the motion along the $x$-axis, is sandwiched by opposite direction streams, which together form two complete circulation cells, and generate a vorticity field. These circulation patterns and central mass accumulation are absent in the classical analysis of the firehose instability. 

\item The bending amplitude, as measured by the curvature of the Laplace plane at the rotation axis and by the isodensity contours distortion, demonstrates that the instability is nonlinear and does significantly affect the motion of stars along the bar, as itemized above.
   
\item The vertical frequency of stellar oscillations in the vicinity of the rotation axis increases with increasing density in the central region, but decreases sharply after buckling. This sudden decrease is associated with a rapid decrease in the vertical acceleration with respect to the Laplace plane. This in turn follows from washing out of the bending which contributed to the vertical acceleration there.   

\item Using the {\it linear} resonances, i.e., ILR and vILR, we find that the vILR appears only with the buckling in the central 100\,pc, is slowly moving out and stagnating at $\sim 6$\,kpc. The planar ILR appears only after buckling at $\sim 2$\,kpc as a single resonance, then splits and the inner ILR moves in to $\sim 1$\,kpc while the outer ILR moves out and ultimately coincides with the vILR. Hence these two resonances overlap with obvious consequences. A substantial region has overlapping ILR and vILR, and this is important.

\item A large fraction of particles that are found in the vicinity of the vertical 2:1 resonance, observed in the $\Omega_{ z}-\Omega_{ x}$ plane, appear to be trapped simultaneously in the planar 2:1 resonance, in the $\Omega_{ R}-\Omega_{ x}$ plane. Note that $\Omega_{ R}$ is a nonlinear extension of $\kappa$, and that $\Omega_{x}$ is defined in the comoving bar frame.  

\item The $\Omega_{z}-\Omega_{x}$ plane reveals the most important aspect in the evolution of this system: most of the particles within the bar are crossing the vertical 2:1 resonance up with the growth of the bar, then crossing the 2:1 line back {\it at the time of the buckling}. This coincidence of the buckling moment with the crossing of the vertical 2:1 resonance serves as the "smoking gun" pointing to the close relationship between both processes --- the nonresonant firehose instability and the resonant excitation of planar orbits. 
    
\end{itemize}  

The main result of this analysis is the overlapping action of the vertical 2:1 resonance with the spontaneous break in the vertical symmetry of the stellar bar. As we have shown, a substantial population of orbits comprising the bar have been swept by the resonance during the short time period of the buckling instability. In other words, the fraction of stellar orbits trapped by the vertical 2:1 resonance is substantial and, therefore, capable of supporting the cohesiveness of the orbital response.

Invoking the firehose instability does not require the resonance action, it predicts the buckling but makes use of somewhat idealized conditions, as we have discussed in Section\,\ref{sec:about}. While the role of the vertical resonance as the triggering mechanism  for the formation of the boxy/peanut shaped bulges, i.e., of bar vertical thickening, has been proposed a long time ago \citep{comb90}, their time coverage of the instability was not sufficient to follow it, and only the final product has been analyzed.  This included the vertical bar thickening, and formation of stellar orbits which explain the peanut/boxy-shaped bulges, the so-called banana (BAN) and anti-banana (ABAN) orbits \citep{pfen91}. 

A subsequent numerical experiment to verify the role of the vILR in the bulge formation proceeded under strict vertical symmetry imposed in simulations \citep{friedli90}. Under these conditions, it was demonstrated indeed that the vILR is responsible for the specific peanut/boxy-shaped bulge formation, but on a secular instead of a dynamical timescale. But what about triggering the buckling process? How is the resonant excitation of orbits related to the symmetry breaking in the bar?

The vertical symmetry of a stellar bar about its midplane of course is never perfect, and is subject to a numerical noise. It is this noise that, if amplified by the vertical resonance, will grow and result in a nonlinear departure from the initial symmetry. If the particle response is individual, and the cohesion or collective response is absent, these particles will populate 3-D orbits on both sides of the midplane. This is associated with the formation of the boxy/peanut shaped bulges which are populated by the particles which increase the energy of vertical oscillations. However, if the particle response is collective, the `macroscopic' departure from the symmetry will follow --- the vertical buckling. 

If the vertical buckling would result from the nonresonant firehose instability, one would not observe the particle trapping by the vertical 2:1 resonance simultaneously with the breaking of the vertical symmetry, and the overlap of the planar and vertical 2:1 resonances would not occur at the same time.

We have demonstrated that the vertical buckling instability affects both the mass distribution and the velocity field in the stellar bars, and the amplitudes of these processes are nonlinear. Our results support the following evolutionary sequence compressed within the time period of buckling, $\sim 0.5$\,Gyr\footnote{The quoted timescale of the buckling instability is just an example. As we show elsewhere, this timescale can vary within a factor of 2--3.}. As the particles approach the 2:1 resonant region on the $\Omega_{ z}-\Omega_{ x}$ plane, the kinetic energy of the motion in the $xy$-plane is pumped into the vertical motions. This decrease in the planar energy results in the slowdown of the particles there. Such a slowdown leads to the increase in the central mass. Increase in the central mass concentration requires increase in $\Omega_{ x}$ along the bar. While increase in the kinetic energy of the vertical motions puffs up the bar in the $xz$-plane, which decreases $\Omega_{ z}$. 

While we cannot rule out completely the contribution from the firehose instability to the buckling process at present, the above sequence leads us to conclusion that the buckling instability is driven mainly by the collective resonant bending of stellar orbits, made possible near the core because the potential is there almost harmonic. And it is the collectiveness of the response which leads to its cohesiveness.  

\section{Conclusions}
\label{sec:conclusions}

In summary, we have analyzed the buckling instability of a stellar bar, focusing on the trigger of this instability. We find that resonant action on the stellar orbits can act to pump the disk energy into vertical oscillations. The substantial overlap between the nonlinear planar and vertical 2:1 resonances which traps a substantial fraction of stellar particles can ensure the cohesiveness of response of stellar orbits and triggers the breaking in the vertical symmetry. While we cannot rule out the contribution of the nonresonant firehose instability to this process, we find strong indications that either the firehose instability has a strong resonance contribution, or the resonances act as a sole trigger for buckling instability. 

What remains to be further examined is the uniqueness of the process that leads the bar towards peculiar resonant conditions of vertical instability. In other words, is the secondary buckling \citep{marti06} triggered by the resonances as well?  Moreover, does the buckling instability in the bar immediately follow the bar instability, or these two instabilities can be well separated in time --- we address this issue elsewhere. 


\section*{Acknowledgements}

We thank Phil Hopkins for providing us with the latest version of the code and Angela Collier for sharing some of the analysis software.   I.S. is grateful for a generous support from the International Joint Research Promotion Program at Osaka University. This work has been partially supported by the Hubble Theory grant HST-AR-18584, and by JSPS KAKENHI grant 16H02163 (to I.S.). The STScI is operated by the AURA, Inc., under NASA contract NAS5-26555.  Simulations have been performed using the University of Kentucky Lipscomb Computing Cluster. We are grateful for help by Vikram Gazula at the Center for Computational Studies of the University of Kentucky.

\section*{Data Availability}

The data presented in this work can be obtained upon reasonable request.



\bibliographystyle{mnras}
\bibliography{paper} 

\appendix
 
\section{NONLINEAR ORBIT CHARACTERIZATION}
\label{sec:append}
 
\begin{figure*}
\center 
	\includegraphics[width=0.9\textwidth]{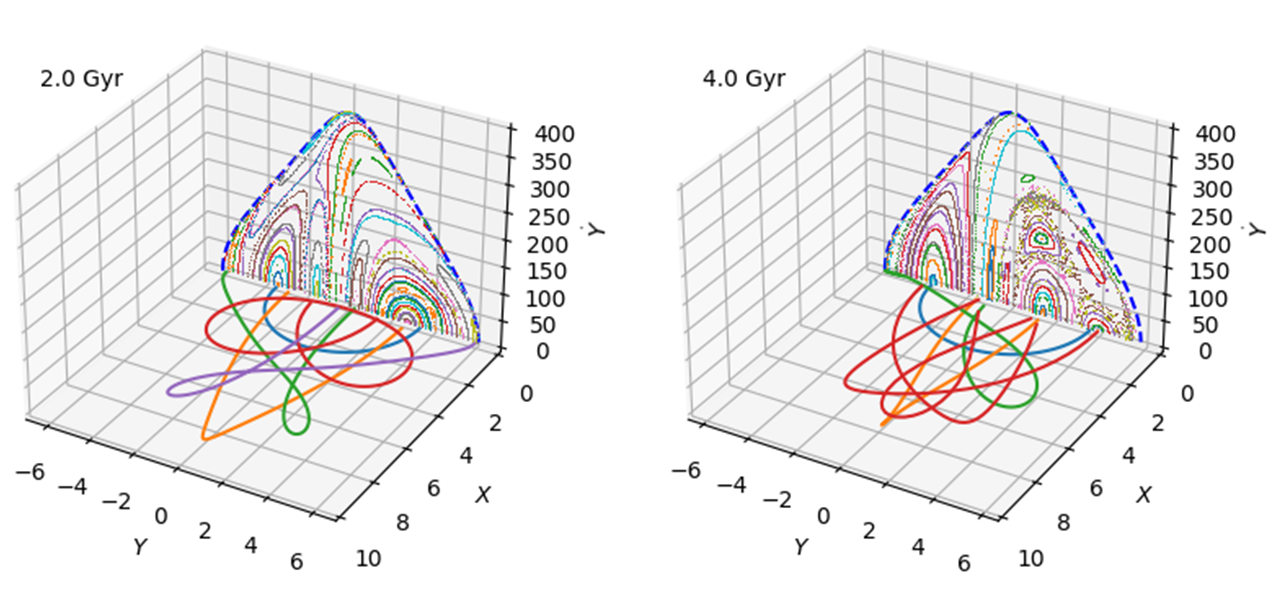}
	\caption{Surface-of-Section with corresponding planar periodic orbits. At $t= 2$\,Gyr (left frame), the $x_1$ (yellow), $x_4$ (blue), 4/1 (red) families, along with a pair of 3/1 orbits (green and purple) are present. Only the $x > 0$ parts of the orbits are shown. The X and Y coordinates are in kpc and $\dot Y$ are in ${\rm km\, s^{-1}}$. No $x_2$/$x_3$ orbits are found in this model. Later, at $t = 4$\,Gyr (right), we find again the $x_1$ and $x_4$ orbits, the 3/1 with a loop at the end of the y-axis (green), along with several multi-periodic families, including a 3-periodic (red). The increase in stochasticity is noticeable in the outer bar region. }
    \label{fig:sos}
    \end{figure*}
 
In order to characterize the orbital structure we have carried out a nonlinear periodic orbit analysis. For this we have fit a recursive triaxial functional form to the potential at two different model times with an overall accuracy of 0.7\% within 15\,kpc. As a check we have verified that the Laplacian of the potential is positive over the fitted domain.

In Figure\,A1 we show the Surface-of-Section with corresponding planar periodic orbits before the buckling at $t = 2$\,Gyr (left) and at $t = 4$\,Gyr (right). Before buckling we find the usual barred potential with the $x_1$, $x_2$ and 4/1 families, along with a pair of 3/1 (triangular) families. No inner Lindblad families are found in this model at any time. After buckling there is in addition several multi-periodic families accompanied by an increase of stochastic regions in the outer bar. 

\begin{figure*}
\center 
	\includegraphics[width=0.9\textwidth]{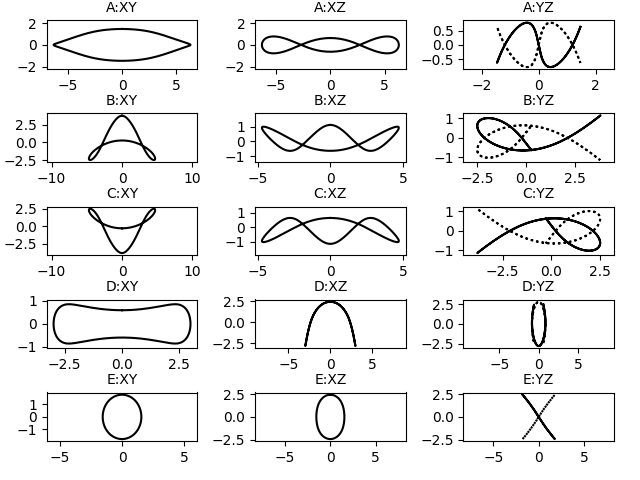}
	\caption{Periodic 3D families.  From top-to-bottom are (A) the main family supporting the bar, (B and C) a pair of 3/1, and unstable (D) BAN like and (E) anomalous retrograde families. The coordinates are in kpc.}
    \label{fig:orbits}
    \end{figure*}  

The main 3D families in the model are shown in Figure\,A2 in three different projections.  Each frame has a pair of orbits (solid and dashed) that reflect the $z$ symmetry of the potential.  The main family supporting the bar shown in the first row (A) is a 1:1:2 ($\Omega_x:\Omega_y:\Omega_z$) orbit that has the characteristic x1 shape in the x-y projection.  The next two rows (B and C) show a pair of 1:1:3 orbits, while the last two rows (D and E) show unstable 1:1:2 and 1:1:1 orbits, respectively.
 

\bsp	

\label{lastpage}
\end{document}